\documentclass[format=acmsmall,authorversion, screen]{acmart}

\usepackage{booktabs} 
\usepackage[inline]{enumitem}
\usepackage[ruled]{algorithm2e}
\usepackage{tabularx}
\usepackage{multirow}
\usepackage{soul}
\usepackage{enumitem}
\newcommand{\subscript}[2]{$#1 _ #2$}
\usepackage{color, colortbl}
\usepackage[markup=underlined]{changes}

\setcopyright{acmlicensed}

\acmISBN{}

\acmJournal{TCPS}
\acmYear{2020} 
\acmVolume{1} 
\acmNumber{1} 
\acmArticle{1} 
\acmMonth{1} 
\acmPrice{15.00}
\acmDOI{10.1145/3418528}

\definecolor{Gray}{rgb}{0.5, 0.5, 0.5}

\begin{document}
\title{Control Communication Co-Design for Wide Area Cyber-Physical Systems}

\author{Laksh Bhatia}
\affiliation{%
  \institution{Imperial College London}
  \streetaddress{South Kensington Campus}
  \city{London}
  \state{UK}
  \postcode{SW7 2AZ}
}
\email{laksh.bhatia16@imperial.ac.uk}

\author{Ivana Tomi\'c}
\affiliation{%
  \institution{Imperial College London}
  \streetaddress{South Kensington Campus}
  \city{London}
  \state{UK}
  \postcode{SW7 2AZ}
}
\email{i.tomic@greenwich.ac.uk}

\author{Anqi Fu}
\affiliation{%
  \institution{Imperial College London}
  \streetaddress{South Kensington Campus}
  \city{London}
  \state{UK}
  \postcode{SW7 2AZ}
}
\email{a.fu@imperial.ac.uk}

\author{Michael Breza}
\affiliation{%
  \institution{Imperial College London}
  \streetaddress{South Kensington Campus}
  \city{London}
  \state{UK}
  \postcode{SW7 2AZ}
}
\email{michael.breza04@imperial.ac.uk}

\author{Julie A. McCann}
\affiliation{%
\institution{Imperial College London}
  \streetaddress{South Kensington Campus}
  \city{London}
  \state{UK}
  \postcode{SW7 2AZ}
}
\email{j.mccann@imperial.ac.uk}
\authorsaddresses{Authors' addresses: L.Bhatia, I.Tomi\'c, A.Fu, M.Breza and J.McCann, Imperial College London, South Kensington Campus, London, SW7 2AZ, United Kingdom; emails:laksh.bhatia16@imperial.ac.uk, i.tomic@greenwich.ac.uk, a.fu@imperial.ac.uk, michael.breza04@imperial.ac.uk, j.mccann@imperial.ac.uk}

\renewcommand{\shortauthors}{L. Bhatia et al.}

\begin{abstract}
Wide Area Cyber-Physical Systems (WA-CPSs) are a class of control systems that 
integrate low-powered sensors, heterogeneous actuators and computer 
controllers into large infrastructure that span multi-kilometre distances.
Current wireless communication technologies are incapable of meeting the communication requirements
of range and bounded delays needed for the control of WA-CPSs.
To solve this problem, we use a Control-Communication Co-design approach for WA-CPSs, that we refer to as the $C^3$ approach, to design a novel 
Low-Power Wide Area (LPWA) MAC protocol called \textit{Ctrl-MAC} and 
its associated event-triggered controller that can guarantee the closed-loop stability of a WA-CPS.
This is the first paper to show that LPWA wireless 
communication technologies can support the control of WA-CPSs. LPWA technologies are
designed to support one-way communication for monitoring and are not 
appropriate for control. 
We present this work using an example of a water distribution network application 
which we evaluate both through a co-simulator (modelling both physical and cyber
subsystems) and testbed deployments. 
Our evaluation demonstrates full control stability, with up to $50$\%
better packet delivery ratios and $80$\% less average end-to-end delays when compared to
a state of the art LPWA technology. 
We also evaluate our scheme against an idealised,
wired, centralised, control architecture and show that the controller 
maintains stability and the overshoots
remain within bounds.
\end{abstract}
%
%



\maketitle

\section{Introduction} \label{sec:intro}

\emph{Wide Area Cyber-Physical Systems (WA-CPSs)} are a confluence of emerging 
communication and
control technologies.WA-CPSs enable the reliable operation of large scale infrastructure 
such as solar farms, precision agriculture, offshore oil facilities, gas pipelines, 
and water/waste distribution networks. All of these examples are geographically 
distributed beyond the size of a single building.
For the most part, dynamics of these
systems are relatively slow; pumps and motors can have long start-up times, a change
in valve levels can take a few seconds \cite{actuators}.
The challenge of engineering a WA-CPSs is the tight coupling between the requirements 
of the physical system's controller and the sensor/actuator/controller communication system. 
This relationship is typical of all control systems. We argue in this work that if we embed 
this coupling in a deeper way than just understanding abstract notions of delay, i.e. we 
inform the parameters of the communication system directly, then we can achieve what was hitherto 
not readily possible.

Low Power Wide Area (LPWA) technologies \cite{Raza2017} create the potential for the development of WA-CPSs. They 
feature low energy consumption and long-range wireless 
communication. The problem with the use of current LPWA technologies in control applications is 
that they have been designed for monitoring; supporting sensor to gateway traffic only. Control applications also require 
communication from the gateway to the actuator. Current LPWA technologies have large non-deterministic communication delays, which makes stabilising a control system problematic \cite{Tomic2018}.

Communication for smaller scale controlled systems commonly uses wired connections between the sensors, 
actuators and the controller. The cost of wired communication increases with the size of the system. A wireless approach 
would reduce the cost of the wires, and lower the cost to install, maintain and scale as the system grows. 
Current wireless technologies like WirelessHART, ISA 100.11a, IEEE 802.11 are incapable of meeting the \textit{communication
requirements} of range and bounded communication delays needed for the control of \emph{large area} infrastructure.
Industrial wireless control protocols, such as WirelessHART
\cite{Song2008}, only support control for ranges up to 100m. Multi-hop network protocols, such as 6TiSCH, 
can be used to extend communication range at the cost of reliability \cite{Ilajosana2019}. The lack of reliability manifests itself as
non-deterministic complexity and delays making WirelessHART and 6TiSCH unsuitable for WA-CPS \cite{Lu2016}. 
LPWA 
technologies can solve the problem of unreliable wireless communication over distance and are a candidate solution to the WA-CPS challenge - but with some adjustments.

The \textit{control requirements} of large scale infrastructure are similar to current control systems that use supervisory control and data acquisition (SCADA). Both need to guarantee system stability.
SCADA systems accommodate continuous or periodic sensing to exchange potentially large volumes of data
between the sensors, controller and actuators at a constant rate. This approach to 
control is energy inefficient and requires high communication bandwidth and reliability.
The requirement of energy-efficient, low bandwidth sensing for control has led to a new control 
theory approach called Event Triggered Control (ETC). In ETC the sensors transmit readings only 
when required, not constantly or with a fixed period as in SCADA systems. ETC schemes have been 
designed and tested with success on small scale systems like the Double tank system
\cite{araujo2014system} and the WaterBox \cite{kartakis2017communication}. ETC is a strong candidate 
to enable control over WA-CPS using low-powered sensor systems. But, ETC
has yet to be evaluated for larger-scale systems where the sampling intervals are larger, and the delays
can be longer than for Double tank system or WaterBox. 

In this paper, we employ a \textit{Control Communication Co-design 
approach} which we refer to as the $C^3$ approach,
a novel co-design approach
for the control and communication systems at design-time and during run-time. At design time, we take into account the effect of the control and communication parameters on the performance of each 
other. We determine these parameters to ensure that the system is globally exponentially stable when used at run-time and do not require further updating. At run-time, the gateway schedules communication access to the sensors only upon need. The $C^3$ approach reduces delays, improves, 
reliability and enables the system to scale beyond the current state-of-the-art LoRaWAN.

A co-design approach is essential to be able 
to guarantee the robustness and fault-tolerance of a WA-CPS
because of the influence of the communication system and control system upon each other \cite{8166737}. 
The core of our approach is that we specify the control and communication parameters
and define system constraints specifically for WA-CPS.  
We use the $C^3$ approach to engineer a WA-CPS distributed feedback control system which depends upon LPWA 
communication technology and solves the challenges arising from the tight
coupling between communication reliability and its effect on the 
stability guarantees of the controller.
As an outcome of the $C^3$ approach, we present \textit{Ctrl-MAC}, a new link-layer 
protocol and its associated event-based control model. \textit{Ctrl-MAC} 
addresses the imperfections of single-hop LPWA networks, not designed for control purposes,
in terms of delay bounds, message loss, two-way traffic and duty cycling. The associated event-based
control model and its parameters are the first to demonstrate a provable, workable solution for control
in WA-CPS using resource constrained sensor and actuator nodes.

We evaluate \textit{Ctrl-MAC} and its controller, with a simulated 
large-scale water supply network parameterised with real-world data, and in controlled
deployments. 
Our $C^3$ approach is purposefully generic, supporting many WA-CPS 
application classes over most current LPWA network technologies. 
In this work we use LoRa, a LPWA technology, to implement
\textit{Ctrl-MAC}.
\vspace{4pt}
\\
\textbf{Contributions.} We provide the following contributions:
\begin{itemize}
    \item We are the \textit{first to practically demonstrate that LPWA technologies 
    like LoRa can satisfy the requirements of WA-CPS control}. Using our $C^{3}$ 
    approach, we design {Ctrl-MAC} and the associated controller. \textit{Ctrl-MAC} is
    the first to provide a backbone for guaranteed simultaneous control of multiple physical processes that require bidirectional data communication. \textit{Ctrl-MAC} also accounts
    for aperiodic and bursty traffic patterns. The performance of \textit{Ctrl-MAC} achieves up to $50$\%
    better packet delivery ratio and five times lower average round trip times than
    LoRaWAN \cite{LoRaWAN2015}, a LPWA MAC protocol. We also compare our solution to an ideal, wired, centralised control architecture and show that our controller can maintain stability and the overshoots remain within bounds. The results are validated by
    a test-bed deployment which shows that our system maintains performance and
    operational bounds. 
    \item  We extend the current state-of-art research in ETC systems for large scale 
    geographically distributed system. By extending our existing work on general 
    event-based control, we derive a \textit{control model, formally analysed to 
    guarantee the closed-loop stability of large-scale water distribution scenarios}.
    \item To evaluate our approach, we built a \textit{cyber-physical co-simulation environment} that connects Matlab/Simulink and OMNeT++. We could find no other simulation environment that could fully model a LoRa network while also providing an accurate water process model. The lack of a suitable simulation environment motivated us to produce a new one to evaluate our work. 
\end{itemize} 

\vspace{4pt}

\textbf{Roadmap.} Sec.~\ref{sec:sys_overview} presents the system architecture and defines the co-design constraints. Sec.~\ref{sec:main_model} presents \textit{Ctrl-MAC} which enables the stable operation of the large-scale control system presented in Sec.~\ref{sec:water_sys_control}. Sec.~\ref{sec:evaluation} presents the evaluation of \textit{Ctrl-MAC} and its controller via a co-simulation environment and a real testbed. Sec.~\ref{sec:related_work} gives an overview of the related work. Sec.~\ref{sec:limitations} discusses the limitations of our approach and we conclude the paper in Sec.~\ref{sec:conclusion}.

\section{System Overview} \label{sec:sys_overview}

In this section, we explain our WA-CPS specific control communication 
co-design approach $C^{3}$. 
We first present the architecture and a model of WA-CPSs. 
Then we describe the set of key design parameters from the 
control system and the communication system that we map into the 
co-design constraints.

\subsection{System Architecture} \label{sec:sys_arch}

\begin{figure}[!t]
\begin{center}
\includegraphics*[width=0.5\linewidth]{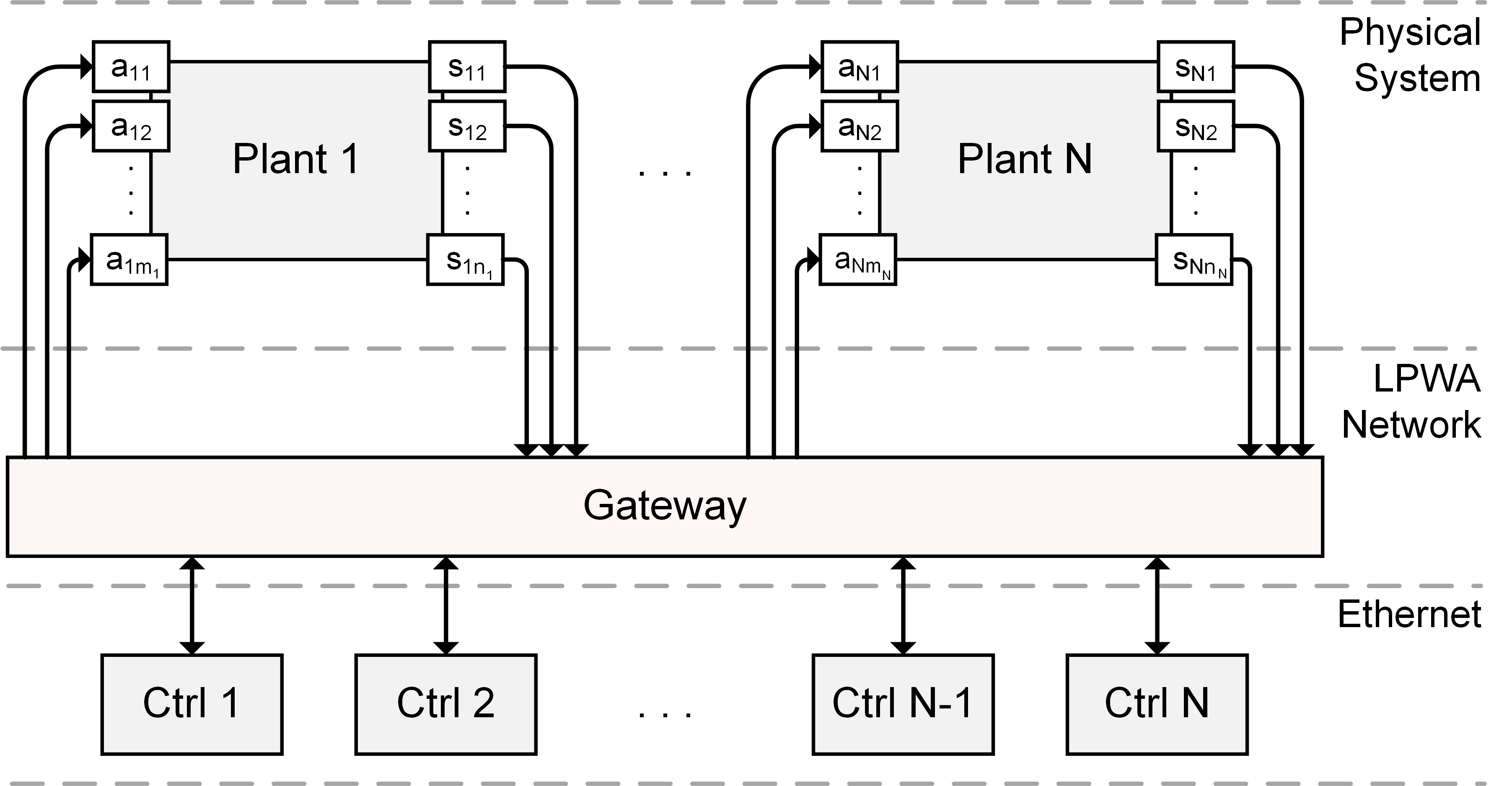}
\caption{\label{fig:architecture} The architecture of WA-CPS.}
\vspace{-5mm}
\end{center}
\end{figure}

Our $C^{3}$ approach focuses on WA-CPS with an architecture such as that shown in Fig.~\ref{fig:architecture}.\vspace{4pt}\\
%
\textbf{Physical System and Controller.}
A WA-CPS physical system consists of $N$ subsystems each labelled $i$,
where $i \in \{1, \ldots, N\}$. Each subsystem consists of a single plant and a single
controller, and both are labelled with the same $i$ as their subsystem.
We describe the $i$th subsystem's plant using a linear time-invariant model:
\begin{equation}
\label{dpetc:eq:plant}
\dot{\xi_{i}}(t)=A_{i}\xi_{i}(t)+B_{i}v_{i}(t).
\end{equation}
The vectors $\xi_{i}(t)\in\mathrm{R}^{n_i}$ and $v_{i}(t)\in\mathrm{R}^{m_i}$
are the state and control input vectors of subsystem $i$,
respectively.
$n_i$ and $m_i$ correspond to the number of sensors and actuators of the subsystem $i$, respectively.
$A_i$ and $B_i$ are the matrices appropriate to the application.

The plant $i$ is instrumented with sensors $\{s_{i1}, s_{i2}, \ldots, s_{in_i}\}$ to measure
its physical processes (individual states of 
$\xi_{i}(t)$). The controller receives these measurements
and processes them according to an underlying control scheme. 
The goal of the controller is to maintain the values in vector $\xi_{i}(t)$ in Eq.~\ref{dpetc:eq:plant}
at a certain fixed value, a reference or set-point.
We use a linear input-to-state feedback controller given as:
\begin{equation}
\label{dpetc:eq:controller1}
v_{i}(t)=K_{i}\hat{\xi_{i}}(t),
\end{equation}
where $\hat{\xi_{i}}(t)\in\mathrm{R}^{n_i}$ are the measured values of states transmitted
to the controller. $K_{i}$ is a designed control gain for which the closed-loop
matrix $A_{i}+B_{i}K_{i}$ is a Hurwitz matrix.

The output of the controller is an action that is sent to actuators $\{a_{i1},
a_{i2}, \ldots, a_{im_i}\}$ to influence the dynamics of the plant. The communication
between the plant and the controller forms a single feedback control loop and 
spans potentially kilometres. 

The feedback control loops in WA-CPS systems have dynamics where changes occur on the order of tens of seconds rather than milliseconds. We use a two-step process to engineer a controller
that can achieve and maintain the stable operation of a subsystem.
First, we design the control gain $K_i$ in Eq.~\ref{dpetc:eq:controller1} 
so that the controller has a slow
response to system changes. The slow response 
reduces the overshoot (deviation from the reference or 
set-point) of the system's state. Second, we choose an event-based control system because it minimises
the overall number of data transmissions, saving bandwidth, which are requirements
of our LPWA networks. We present the definition of event-based
control in Sec.~\ref{sec:water_sys_control}. The set of parameters
that guarantee the stability of the subsystem $i$ (based on the Corollary~\ref{dpetc:corollary:stabilitywithdelay}
in Sec.~\ref{sec:stab_anal}) are:
\begin{itemize}
    \item \textit{Sampling interval, $h_i$} - The time 
between individual samples. We address systems whose dynamics change slowly. The 
sampling time supported by the controller
is in the order of tens of 
seconds (from $1$ to $50$ seconds).
    \item \textit{Maximum allowable transmission delay, ${\tau_d}_i$} - 
The maximum interval between sensor node state measurement and the 
time the actuator nodes receive updated controller actions. The delay supported by
the $C^{3}$ approach is in the order of seconds (from $0$ to $15$ seconds).
    \item \textit{Event trigger parameter, $\sigma_i$} - It defines the
size of the error between the estimated and measured states. A small $\sigma_i$ is more
sensitive to change and triggers new control inputs at a faster rate which leads to a
larger number of events. We chose to use the largest $\sigma_i$ that can still
guarantee stable system operation to minimise energy consumption.
    \item \textit{Convergence rate of the control system, $\rho_i$} - WA-CPSs 
do not have strict restrictions on convergence speed as long as they can meet the delay tolerances,
hence $\rho_i=0.001$ for all $i$.
\end{itemize}
\vspace{2pt}
These parameters are core to our $C^{3}$ approach as they are shared with and influence the communication system behaviours described next.

In this work, we assume that sensor nodes and actuator nodes are not collocated. We assume that all of the sensors are low-power embedded devices and battery-powered (either
primary or secondary batteries). Actuators are battery or mains powered when
the actuation function requires relatively high power.
However, our approach is directly applicable to wireless networks where the sensors and the
actuators harvest energy for energy-neutral operation, which is relatively deterministic and appropriate
\cite{dicken2012power}. \vspace{4pt}\\
\textbf{Wireless Communication System.}
We focus on LPWA networks because they are designed as low-power one-hop wireless 
solutions for long-distance communication. LPWANs enable communication between the 
sensors, the actuators, and the controller over multi-kilometre distances. Single-hop wireless networks
are better than multi-hop wireless networks for control-based systems due to their
increased reliability and lower (easier to determine) communication delays \cite{Lu2016, Ilajosana2019}.

All $N$ subsystems in Fig.~\ref{fig:architecture}, and therefore $N$ feedback control loops, are coupled through
the shared LPWA communication network and use a single gateway. 
We use the term gateway and controller interchangeably and treat them as one device.
The gateway and the controllers communicate
via wired communication which is instantaneous and reliable. In practice, the 
controller and gateway may reside on different devices, the controller 
on a server along with other processes (logging and management) and the gateway 
in a position favourable to receive radio communication (on a rooftop).

We assume that packet loss occurs only as a result of message collisions when two 
packets arrive at the same time. Sensors always send up-to-date information {and}
retransmit if no acknowledgement is received. If information is delayed due to 
retransmission and a new message is generated, the original information is replaced 
by the most up-to-date information.

We specify the parameters of the LPWA network that affect 
the function of the network and the controller:
\begin{itemize}
    \item \textit{Channel access time, $t_{ca}$} - The maximum time that a sensor node takes
    to access the channel to send, dependant on the channel access scheme used.
    The goal of our $C^{3}$ approach is to minimise $t_{ca}$ and any other waiting times (delays) due to the
    time-criticality of information in controlled systems.
    \item \textit{Time to complete the transmission, $t_t$} - The maximum time that it will
    take for a sensor node to transmit its data to the controller (assuming it
    accessed the channel beforehand) and for the controller to update and transmit
    its action to an actuator node. We assume reliable communication, so a lost packet
    triggers a resend. If newer information is available on the sensor node, then the old
    information is replaced by the newer information for sending.
    \item \textit{Duty cycle, $DC$} - The maximum percentage of time spent in communication
    per node per channel for a given time interval. For instance, a 1\% duty cycle allows a sensor node
    to send for 1 second out of every 100 seconds. We assume that any communication network
    that we use will have a fair use policy enforced on the system to allow multiple
    users.
\end{itemize}

In the next section we will discuss how the key parameters for the control system and 
communication system influence each other. The design of the system by the selection 
of their values is the core of our $C^{3}$ approach.

\subsection{Co-design Constraints} \label{subseq:design_goals}

$C^{3}$ maps an event-based control scheme to a LPWA protocol, and vice versa, through their
coupled parameters. The set of key design parameters from the event-based controller
is $\{h_i, {\tau_d}_i, \sigma_i, \rho_i \}$ and the parameters for the communication are
$\{t_{ca}, t_{t}, DC\}$. The coupling between two sets of parameters is twofold and given via two
co-design constraints. These are:
\begin{enumerate}[label=\subscript{\textbf{C}}{{\textbf{\arabic*}}}]
    \item The control parameter ${\tau_d}_i$ is directly affected by the communication parameters:
    $t_{ca}$, $t_t$ and $DC$. The stability can be guaranteed (based on 
    Corollary~\ref{dpetc:corollary:stabilitywithdelay} in Sec.~\ref{sec:stab_anal}) only if the service
    time for each event is less than ${\tau_d}_i$ ($0$ to $15$ seconds) while complying with the $DC$ restriction. We define the service time as the sum of delays due to $t_{ca}$ and $t_t$.
    \item The choice of control parameters $h_i$, $\sigma_i$ and $\rho_i$ determine the
    number of events which cause transmissions. These constitute the load on the communication 
    network. The network load cannot be larger than the maximum network capacity.
    Again, the choice of these parameters is dependant upon \textbf{C\textsubscript{1}}, and stability
    has to be guaranteed (based on Corollary~\ref{dpetc:corollary:stabilitywithdelay} in Sec.~\ref{sec:stab_anal}).
\end{enumerate}

Our $C^{3}$ approach focuses upon these co-design constraints to enable the design of the WA-CPS.
The next two sections will focus on the two sides of our $C^{3}$ approach separately. First, we focus
upon the design of a communication protocol for WA-CPSs, then we look at the design of a controller
for WA-CPSs. We show how the use of $C^{3}$ ensures that the resulting communication and
control systems meet the above co-design constraints and guarantee stability.

\section{An LPWA Communication Infrastructure for Control} \label{sec:main_model}

In this section, we present a novel LPWA MAC protocol called 
\textit{Ctrl-MAC}. \textit{Ctrl-MAC} is the result of the 
application of our $C^{3}$ approach and is agnostic to specific LPWA 
technologies such as LoRa, NB-IoT or Sigfox \cite{Raza2017}. First, we present the
operating principle and the implementation details of \textit{Ctrl-MAC}.
Then, we present a delay analysis to show that \textit{Ctrl-MAC} bounds 
communication delays and adds guarantees to the low data rates and unreliable
communication typical of all of the current, major LPWA technologies. In doing this,
\textit{Ctrl-MAC} ensures that the communication requirement of control
systems (the co-design constraint \textbf{C\textsubscript{1}}) is met. 
Finally, we evaluate \textit{Ctrl-MAC} against LoRaWAN, a state-of-art LPWA
MAC protocol. Our results confirm that \textit{Ctrl-MAC} achieves lower,
bounded communication delays, reliable two-way communication and higher throughput
when compared to LoRaWAN.

\subsection{\textit{\textit{Ctrl-MAC}} Operating Principle}

\textit{\textit{Ctrl-MAC}} operates in two phases: 
\begin{enumerate*}
\item Sensing and data transmission phase
\item Control update and actuation phase.
\end{enumerate*}
These are given next.
\vspace{4pt}\\
\textbf{Phase 1: Sensing and Data Transmission.}
In the sensing and data transmission phase, each sensor
node (within each subsystem) goes through three stages. 

The first stage is when the \textit{sensor node senses the physical
phenomenon}. This data is interrogated to identify if a significant event has occurred. It is transmitted if a significant event is identified. The sensor sleeps until the next sensing period if no significant event is identified.

\begin{figure}[!t]
\begin{center}
\includegraphics*[width=0.4\linewidth]{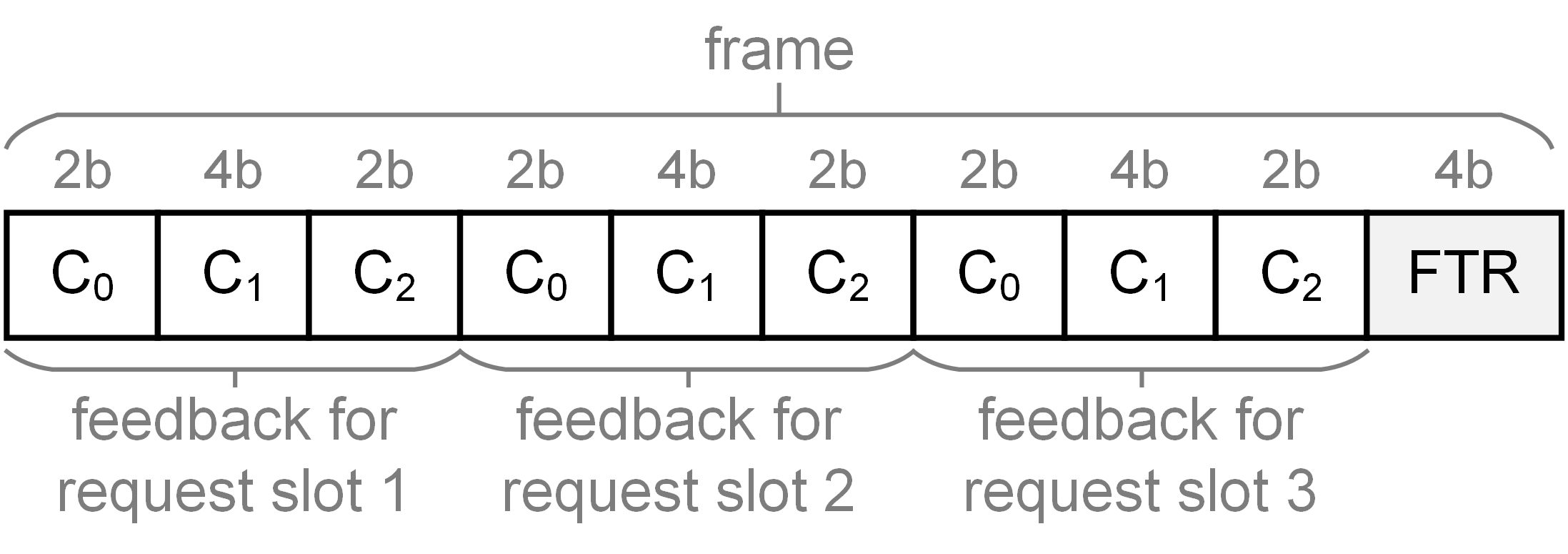}
\caption{\label{fig:message} The RRM structure for $k=3$,
$M_D=3$ which implies that $size(C_2)$=$\log_2{M_D}$, $l=16$.}
\vspace{-5mm}
\end{center}
\end{figure}

Upon the identification of an event, the sensor synchronises
to the start of the request period. The sensor updates its local device clock to the 
start time  of a request reply message (RRM, shown in Fig.~\ref{fig:message}) 
received from the gateway. The gateway sends RRMs periodically. Synchronisation is
required to correct the clock drift caused after the sensor node resumes from 
power-saving deep sleep mode. The sensor node also learns the number of request slots $k$, the duration of each request slot $t_{slot}$ from the RRM.

The second stage is when the \textit{sensor node transmits its data transmission
request}. Once the sensor node is synchronised to the request period, it randomly
chooses which slot to transmit in. The sensor chooses using a uniform random 
distribution of all available request slots instead of a fixed slot or following a
round-robin slot scheduling scheme. This is done to manage the case where there are 
more sensor nodes than available request slots where no fixed schedule
is possible. The sensor makes a data transmission request by transmitting its
ID during its chosen time slot.
If the sensor is the only one requesting a transmission slot, 
its request will be successful; otherwise, there will be a collision (a result of slot
contention) and the request for both sensors will fail.

The sensor node is informed about the success or failure of
its data transmission request in the next RRM. An example 
of a RRM for a system with $k=3$ transmission request 
slots and $l=16$ data slots is shown in Fig.~\ref{fig:message}.
The RRM contains three fields for each transmission request slot. These are:
\begin{enumerate}
    \item The state of the request slot, $C_0 \in \{0,1,2\}$, where $C_0=0$ indicates
    that no sensor node sent a request in this request slot, $C_0=1$ indicates that
    there was no contention in this request slot and $C_0=2$ indicates that a 
    collision occurred in this request slot.
    \item The data slot counter, $C_1 \in \{1, \ldots, l\}$, where $l$ is the number
    of data slots.
    \item The data channel counter, $C_2 \in \{1, \ldots, M_D\}$, where $M_D=M-2$ is
    the number of data channels and $M$ is the overall number of channels.
\end{enumerate}
The last value of an RRM is called the $FTR$. $FTR$ contains
the cumulative sum of all contentions that have not been resolved until now. This includes the current failed and previous
failed transmission requests. The $FTR$ is decremented by $1$ after each RRM.

A sensor node checks the value of $C_0$ in the request slot that it used to
transmit its data transmission request. If $C_0=1$, that sensor node can proceed to stage
three and send data. If $C_0=2$ the data transmission request fails. The sensor node must
return to stage two and send another transmission request at the time defined by the
value of $FTR$. If $FTR>1$, the sensor node counts all of the other request slots that
have $C_0$ equal to $2$. This value is denoted as $r$. The sensor node determines the
position of its request slot compared to other $r-1$ request slots. This value is denoted
as $p$. The sensor node waits for the $(FTR+r-p)$th RRM to retransmit its data transmission
request choosing the slot $k$ randomly.

In stage three, the \textit{sensor node transmits data}. The RRM
contains information to partition data transmissions into $(M-2)$ channels and $l$
data slots. The gateway allocates channel/slot pairs to requesting sensors on a first-come-first-served basis. 
When a sensor node receives an RRM with $C_0=1$ it changes to the channel given in $C_2$ and
transmits its data in the slot given by $C_1$.
If the sensor node senses new data during the time that it is requesting a 
transmission slot, it will send the new data, and discard the old. This is because
the controller requires the newest information to keep the system stable.
\vspace{4pt} \\
\textbf{Phase 2: Control Update and Actuation.}
The second phase of \textit{Ctrl-MAC} is the control and actuation phase.
After the controller receives sampled information from the sensor node, it uses it to 
calculate the control action that should be sent downstream to actuator nodes. This 
constitutes one of the challenges met by our co-design approach. Current LPWA protocols such as LoRa 
and NB-IoT prioritise upstream communication only and are therefore ill-suited to
meet the timing requirements of control systems with feedback loops. Instead,
\textit{Ctrl-MAC} adds bidirectional capability to current LPWA
protocols where the downstream communication is as important as the upstream
communication.

The gateway sends actuation information to the actuator nodes in the downlink
message. These messages are sent periodically so no $DC$ regulations are violated. The size of each downlink message can change depending on how many actuator
nodes need to be updated. The sending time of this message will depend on the size of
the previous downlink message (such that it complies with the DC regulation).

Next, we give the details of the \textit{Ctrl-MAC} implementation and design choices.

\subsection{\textit{Ctrl-MAC} Implementation}
\label{subsec:CTRL_impl}

We implemented \textit{Ctrl-MAC} as a proof of concept on top of LoRa. 
We use LoRa's physical layer because it is robust, has open-source libraries and uses inexpensive
low-powered hardware.
LoRa operates in the unlicensed spectrum and is subject to fair usage policies mandated by legislation. Fair usage takes the form of a self-imposed DC of 1\% and 10\% time-on-air for transmissions when operating in the EU region. As \textit{Ctrl-MAC} is implemented on top of LoRaPHY, 
it is important that we explain our motivation for not using LoRaWAN, a LoRa MAC 
protocol, and the differences between the two.
These are given in Sec.~\ref{subsec_ctrlmac_lorawan}.
Our reference implementation of \textit{Ctrl-MAC} has room for significant 
performance improvement. It purposefully does not exploit any LoRa performance
enhancing features, such as the use of different spreading factors, to be
general to LPWA communication schemes such as NB-IoT or SigFox. Further, in all 
simulations and experiments \textit{Ctrl-MAC} maintains a constant spreading factor of 7
which gives low transmission time
and achieves distances of up to 
$2$km \cite{adelantado2017} in urban areas.

\textit{Ctrl-MAC} assumes that there are four channels available based on the
EU recommendation \cite{LoRaWAN2015}. There are three uplink channels of
$125$kHz with $1$\% DC 
and one downlink channel of $250$kHz with
$10$\% DC. The current LoRaWAN implementation uses only half
($125$kHz) of the allocated downlink bandwidth for the acknowledgements.
\textit{Ctrl-MAC} exploits this feature and uses the remaining $125$kHz to send the
actuation signals to actuator nodes. The actuation signals are periodically transmitted
while complying with the 10\% DC regulation.
The size of the downlink message depends on the number of actuator nodes in the network 
and how many bytes of information are used per an actuator. For example, for
a network with $10$ actuator nodes and $2$Bytes of information with the 10\% DC regulation,
the actuation messages can be transmitted every $0.5$ seconds.

\begin{figure}[!htb]
\begin{center}
\includegraphics*[width=0.93\linewidth]{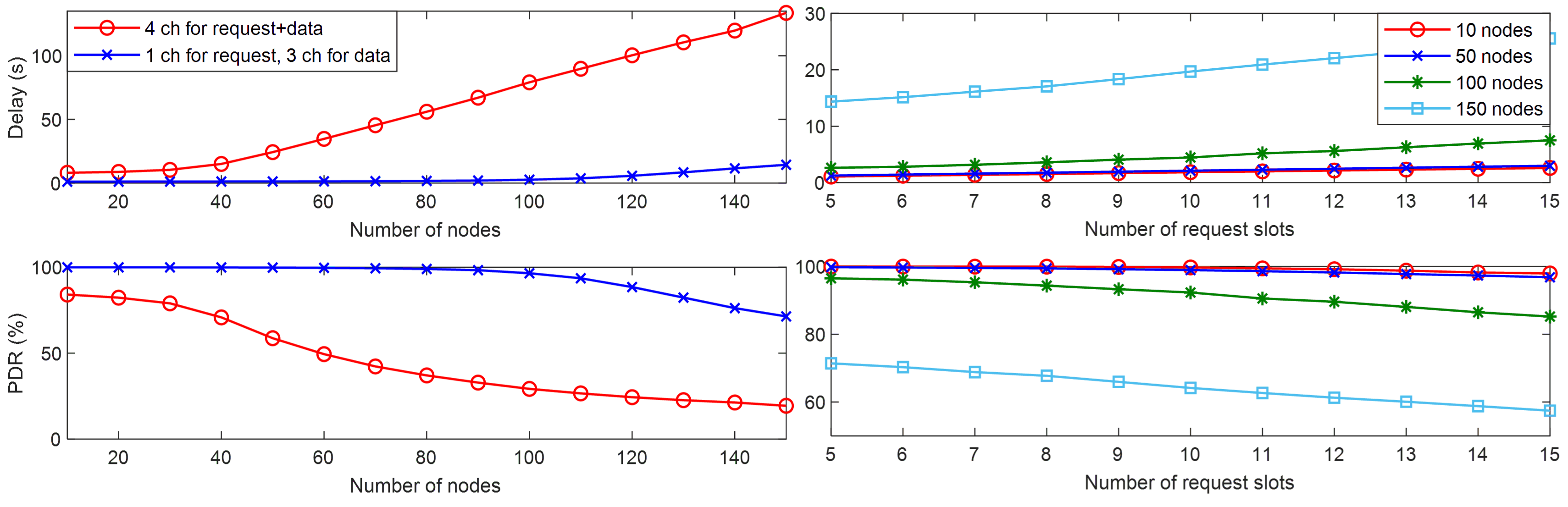}
\caption{\label{fig:MAC_MC_singleC}(left) \textit{Ctrl-MAC} channel disciplines comparison, (right) \textit{Ctrl-MAC} request slots comparison.}
\vspace{-5mm}
\end{center}
\end{figure}

Our implementation uses only a single channel for data request transmissions (with
10\% DC) and the remaining three channels for data transmission.
The results in Fig.~\ref{fig:MAC_MC_singleC} (left) show that when only a single channel is
used for data request transmissions (with 10\% DC) and the remaining three
channels for data transmission, \textit{Ctrl-MAC} achieves lower delays and higher
packet delivery ratios (PDRs) than when all four channels are used for both.
This happens because the data transmission requests and actual data transmissions
operate concurrently unlike the case when the data transmission requests and actual
data transmissions happen on the same channel one after the other. Our implementation
allows more requests to be sent, the contention is resolved faster and data load is distributed
evenly over 3 data channels, and therefore more data can be sent.

\textit{Ctrl-MAC} uses $k=5$ request slots with the duration of $t_{slot}=0.1$ seconds.
The value of $k$ cannot be smaller than $5$ and the request slots can not be of less
time to account for the 10\% DC regulation. The results in Fig.~\ref{fig:MAC_MC_singleC} 
(right) show the degradation of the communication delays and PDRs for all $k$ larger than 5 
and all populations of sensor nodes.

Next, we model \textit{Ctrl-MAC} to analyse its performance in terms of the 
guarantees and achieved communication delays.

\subsection{\textit{Ctrl-MAC} Delay Analysis} \label{sec:delay}

In this section we analyse the communication delays identified by the co-design constraint
\textbf{C\textsubscript{1}} from Sec.~\ref{subseq:design_goals}, using a queue based approach.
We show that the communication delays of \textit{Ctrl-MAC}, due to $\{t_{ca}, t_{t}, DC\}$, are
bounded and do not violate the maximum allowable transmission delay derived from the 
control system. The maximum allowable transmission delay for each subsystem $i$ is
${\tau_d}_i=\tau_d$, which is from $0$ to $15$ seconds. Our analysis obtains the bounds 
of $\{t_{ca}, t_{t}\}$ by deriving the delays of different stages of \textit{Ctrl-MAC},
$\{t_{sync}, t_{req}, t_{send}, t_{update}\}$, where $t_{ca} = t_{sync} + t_{req}$ and 
$t_t = t_{send}+t_{update}$ while complying with the DC regulation. We show that \textit{Ctrl-MAC}
can operate within the communication constraints and keep the total delays less than the maximum
allowable transmission delay (i.e. $t_{sync} + t_{req}+ t_{send}+t_{update}<{\tau_d}$).
\vspace{4pt}\\
\textbf{Sensing and Data Transmission Phase Delay.} 
We model the delays incurred in the three stages of the Sensing and Data Transmission Phase 
and derive their maximum bounds.

The first stage of the first phase is the synchronisation process. This 
occurs when a sensor node has data to send and uses information in the RRM 
received from the gateway. The RRM message is sent periodically and of a fixed size such that
the time on air of the message and its sending rate never violate the DC regulation.
Therefore, the synchronisation process includes a fixed delay, denoted 
as $t_{sync}$, which is in the range $[0,t_{slot} \times k$]. In our system, 
$t_{sync} \in [0,0.5]$ seconds (based on the implementation in
Sec.~\ref{subsec:CTRL_impl} that uses $k=5$ request slots of $t_{slot}=0.1$ seconds).

\begin{figure}[!t]
\begin{center}
\includegraphics*[width=0.5\linewidth]{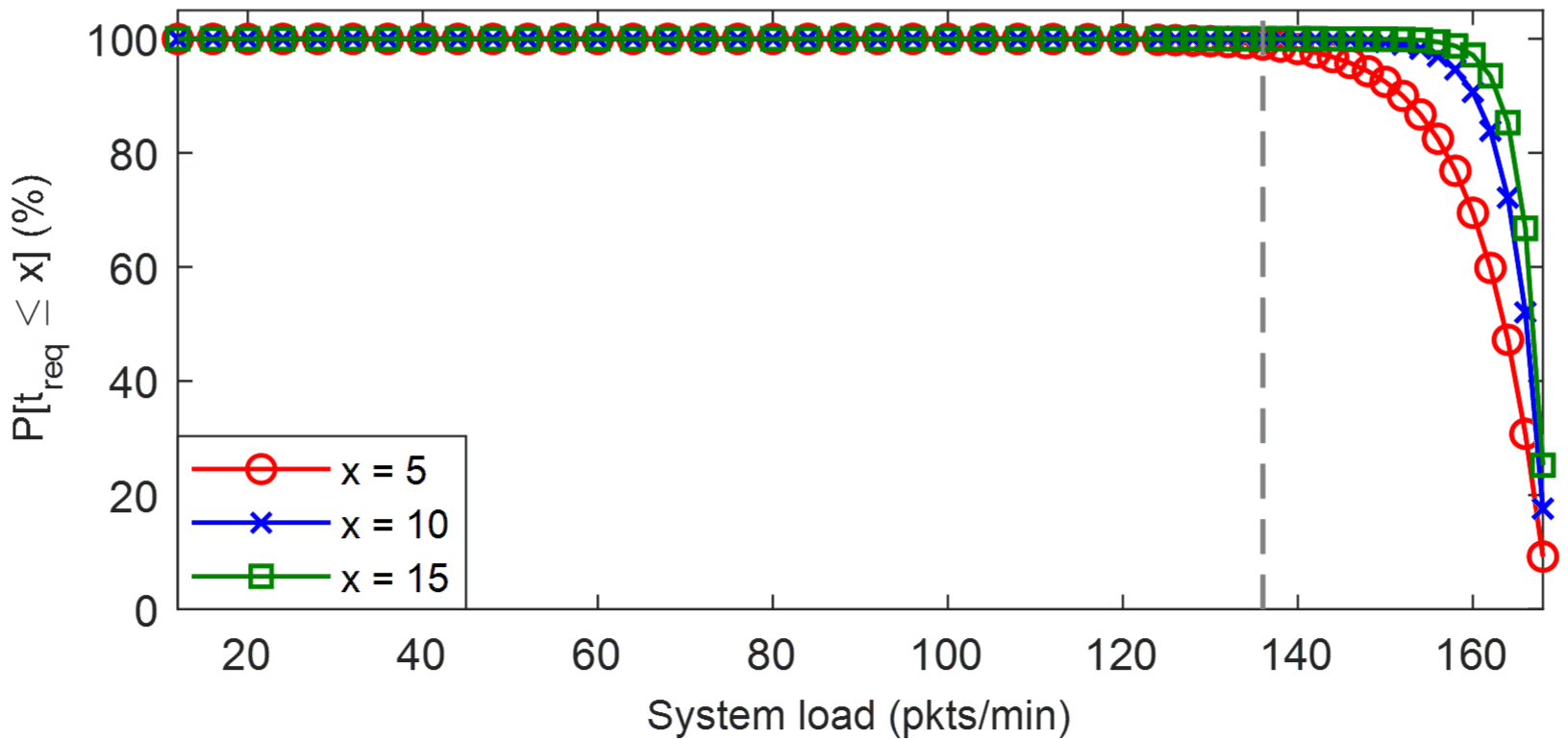}
\caption{\label{fig:probvsload} The probability of achieving $t_{req} \leq x$ for M/M/1 queue.}
\vspace{-5mm}
\end{center}
\end{figure}

After the delay of $t_{sync}$, the sensor node needing to transmit enters the next two stages of phase
one. Delays of the next two stages are denoted as $t_{req}$ and $t_{send}$. Their bounds are analysed
by modelling the two stages as two queues:
\begin{itemize}
    \item \textit{M/M/1 queue, or queue one} - The second stage of the first phase is the data transmission
    request process. We model this process as an M/M/1 queue. We use a queue to analyse the delay, $t_{req}$,
    and its bounds as it allows us to reason about the occurrence of sensor events and the waiting time for a
    sensor node to receive a transmit slot for data communication. We define $t_{req}$ as the mean time for a
    sensor node to successfully request and receive a data slot and is estimated by $1/(\mu_{req}-\lambda)$
    where $\lambda$ is the system load, less than or equal to the number of sensor nodes, and
    $\mu_{req}=ln(1/(1-e^{(-\lambda/k)}))$ is the service rate. We assume that the inter-arrival of sensor
    events and time to get a transmit slot are exponentially distributed (these assumptions are based 
    on the traffic pattern of an event-based control system as presented in Sec.~\ref{subsec:control_operation}). 
    For the system to be stable $\lambda/\mu_{req} < 1$. The delay $t_{req}$ is non-deterministic.
    In the best case scenario, when the sensor node immediately successfully requests and receives a data slot
    $t_{req}=0.2$ seconds. Otherwise, we can analyse that the probability of $t_{req}$ is less than a given
    delay which we denote as $x$. We show through numerical simulation the probability 
    $P[t_{req} \leq x] = 1 - e^{(-(\mu-\lambda)x)}$ for various data loads.
    Results are presented in Fig.~\ref{fig:probvsload}. Our results show a very high probability (more than 
    99\%) of meeting the delay bounds of $5$ seconds for a communication capacity of up to $136$ packets per minute and up to $150$ packets per minute of meeting the delay bounds of $10$ seconds. 
    The packet sizes and time on air of packets at this rate do not violate the DC regulation.
    \item \textit{M/D/$n$ queue, or queue two} - The third stage of phase one is the data transmission stage.
    We model the delay of this stage, $t_{send}$, as an M/D/$n$ queue.
    The term $n$ is the number of data transmission channels available to \textit{Ctrl-MAC}.
    We define $t_{send}$ as the time for a sensor node to send its data to the gateway.
    The queue is deterministic because a sensor node receives the data slot and channel exclusively reserved
    for its transmission from the gateway during the data transmission request stage. The exclusive data 
    slot and channel reservation ensure that the DC regulation is met, and that the time to send is 
    bounded. The delay $t_{send}$ is equal to $1/n+(\lambda_{dt}/(2n\times(n-\lambda_{dt})))$ where
    $\lambda_{dt}$ is the output of the M/M/1 queue given above and the input to
    the M/D/$n$ queue described here. Based on the input rate, $\lambda_{dt}$, is in the range
    of $[12,150]$ packets/minute based on Fig.~\ref{fig:probvsload}, the time to completion is
    $t_{send} \in [0.3,0.45]$.
\end{itemize}
%
\textbf{Control Update and Actuation Phase Delay.} The final delay that we need to consider
is the time for phase two, $t_{update}$, when the controller sends the control input to the
actuator node. The gateway transmits on a separate downlink channel reserved for communication from the
gateway to the actuator. The use of a reserved channel ensures that the $DC$ regulation is met and
that the time of $t_{update}$ is deterministic. The delay is dependant upon the specific LPWAN 
technology and the size of last actuation update sent. In our LoRa implementation, the maximum 
packet size that the controller can send is $222$ bytes. Each actuation update is 2 bytes per 
actuator node, 1 byte for control input and 1 byte for addressing. The maximum number of actuator nodes that 
can be updated with a single packet is $111$. The delay $t_{update}$ is in the range of 
$[0.4,3.6]$ seconds depending on the size of the last actuation update.

Therefore, the total delay introduced by \textit{\textit{Ctrl-MAC}}, $t_{MAC}$ is:
\begin{equation}
    t_{MAC}=t_{sync}+t_{req}+t_{send}+t_{update}.
\end{equation}

According to the co-design constraint \textbf{C\textsubscript{1}} from Sec.~\ref{subseq:design_goals},
the system stability is dependant upon the condition that ${\tau_d} > t_{ca} + t_{t}$ where 
$t_{ca} = t_{sync} + t_{req}$ and $t_t = t_{send} + t_{update}$.
We derived the individual bounds for $t_{sync}$, $t_{req}$, $t_{send}$, and $t_{update}$
while taking the DC regulation into account.
A simple sum of all of our bounds shows that the total delay, $t_{MAC}$ is in the range of
$[0.9,14.55]$ seconds with a very high probability of more than 99\%. The communication delays of
\textit{Ctrl-MAC} do not violate the maximum allowable transmission delay of the system
(i.e. $\tau_d>t_{MAC}^{Upper}$).

\subsection{\textit{Ctrl-MAC} and LoRaWAN} \label{subsec_ctrlmac_lorawan}

In this section we motivate the need for \textit{Ctrl-MAC}, we explain the differences
between \textit{Ctrl-MAC} and LoRaWAN, and show how LoRaWAN is unable to meet communication
requirements for control systems that are reliable communication, two-way communication
and bounded communication delays. We evaluate \textit{Ctrl-MAC} against two versions of LoRaWAN:
\begin{itemize}
    \item \textit{Version one: LoRaWAN} - This is a baseline version that represents the typical
    use case for LPWA technology such as data collection from sensors or smart meters. It exploits
    a pure ALOHA channel access scheme. The sensor nodes are the Class A devices \cite{LoRaWAN2015}
    which use unconfirmed (unacknowledged) messages, i.e. the sensor nodes do not request an
    acknowledgement for their uplink transmissions and hence there is no guarantee
    for successful delivery of the message. The actuator nodes are modelled as Class C devices. As
    mentioned in the LoRaWAN specification \cite{LoRaWAN2015}, Class C devices are always listening on the
    downlink channel unless they are transmitting data. Our actuation messages are sent on this
    downlink channel to the actuators.
    \item \textit{Version two: LoRaWAN++} - This is an improved version of LoRaWAN that enables
    reliable data communication to the gateway. It uses the ALOHA channel access scheme with an
    acknowledgement message from the gateway for every uplink transmission, as described in
    the LoRaWAN specification \cite{LoRaWAN2015}. If no acknowledgement is received, the transmitted
    data is resent up to $8$ times after which it is dropped. Acknowledgements are sent on the
    LoRaWAN downlink channel. In this version we also add a dedicated downlink channel for actuation
    to make this scheme comparable to \textit{Ctrl-MAC}.
\end{itemize}
We evaluate MAC protocol suitability for control applications with three metrics:
\begin{enumerate}
    \item \textit{End-to-End Packet Delivery Ratio} (E2E PDR in \%) - E2E PDR is the ratio of the number
    of unique actuation updates received by the actuator node over the number of unique events generated
    by the corresponding sensor node. This metric indicates the suitability of the MAC protocols for
    control applications in terms of two-way communication. The downstream communication is as important
    as upstream communication.
    \item \textit{End-to-End Delay} (E2E Delay in seconds) - E2E Delay is defined as the
    difference between the time that an actuator node receives the actuation update and the time that
    an event is generated at the sensor node. We assume the computation delays to be zero when
    calculating this metric. This metric indicates the suitability of the MAC protocols for
    control applications in terms of bounded communication delays.
    \item \textit{Uplink Reliability} (UL Reliability in \%) - UL Reliability is the ratio of the 
    number of events successfully acknowledged by the gateway over the number of events successfully
    delivered at the actuator nodes. This metric gives the guarantees that the controller has complete
    and up-to-date information to maintain system stability. For example, a 10\% uplink reliability would
    indicate that only 10\% of the delivered messages from sensor nodes were reliably acknowledged.
    The sensor nodes may or may not have information about the remaining 90\% which introduces
    uncertainties in the system.
\end{enumerate}

We compare the E2E PDR, E2E Delay and UL Reliability of \textit{Ctrl-MAC}, LoRaWAN and LoRaWAN++
assuming four channels, a varied number of nodes (${10,50,100,150,200}$) and two transmission
patterns. First, we chose constant periodic transmissions to represent the worst-case scenario
of the event-triggered control (i.e. at every sampling period, the transmission is triggered -
in reality there would be periods of no events). Each sensor periodically sends data with an interval of ${10, 30, 50}$ seconds. Second, we chose exponential transmissions with a mean of
${10, 30, 50}$ seconds to represent aperiodic and bursty traffic patterns specific to
the event-triggered control. This range of sending rates (inter-arrival times) corresponds to the
sampling interval of slow-loop control systems as it is shown in Sec.~\ref{subsec:control_operation}.
Results are presented in Tables~\ref{tab:pdr_distr}, \ref{tab:delays_distr} and \ref{tab:UL_rel}.
\begin{table}[!t]
\small
\setlength{\tabcolsep}{3pt}
\caption{E2E PDR (\%) of \textit{Ctrl-MAC}/LoRaWAN/LoRaWAN++ under different data loads and data distributions.}
\vspace{-3mm}
\renewcommand{\arraystretch}{0.8}
\begin{center}
\begin{tabularx}{1\textwidth}{l | c c c c c | c c c c c | c c c c c}
\toprule
& \multicolumn{5}{c|}{\textbf{\textit{Ctrl}-MAC}} & \multicolumn{5}{c|}{\textbf{LoRaWAN}} & \multicolumn{5}{c}{\textbf{LoRaWAN++}}\\
\midrule
\textbf{Size} & \textbf{10} & \textbf{50} & \textbf{100} & \textbf{150} & \textbf{200} & \textbf{10} & \textbf{50} & \textbf{100} & \textbf{150} & \textbf{200} & \textbf{10} & \textbf{50} & \textbf{100} & \textbf{150} & \textbf{200} \\
\midrule
\textbf{P(10s)} & \cellcolor{Gray!20}99.99 & 67.72 & 33.62 & 22.34 & 16.86 & \cellcolor{Gray!20}99.98 & 64.32 & 39.82 & 30.26 & 23.22 & 80.85 & 40.31 & 31.39 & 25.89 & 20.24 \\
\textbf{P(30s)} & \cellcolor{Gray!20}99.99 & \cellcolor{Gray!20}99.99 & \cellcolor{Gray!20}98.23 & 67.20 & 50.77 & \cellcolor{Gray!20}99.98 & 80.23 & 72.64 & 57.6 & 48.14 & \cellcolor{Gray!20}99.96 & 86.71 & 40.11 & 27.56 & 21.04 \\
\textbf{P(50s)} & \cellcolor{Gray!20}99.99 & \cellcolor{Gray!20}99.99 & \cellcolor{Gray!20}99.99 & \cellcolor{Gray!20}99.99 & 83.74 & \cellcolor{Gray!20}99.99 & \cellcolor{Gray!20}93.13 & 85.56 & 65.18 & 62.95 & \cellcolor{Gray!20}99.99 & \cellcolor{Gray!20}93.06 & 63.37 & 45.31 & 34.50 \\
\textbf{E(10s)} & \cellcolor{Gray!20}99.98 & \cellcolor{Gray!20}97.59 & 52.73 & 35.27 & 26.43 & 84.42 & 66.44 & 51.32 & 41.45 & 33.44 & 86.31 & 38.54 & 21.34 & 14.66 & 10.94 \\
\textbf{E(30s)} & \cellcolor{Gray!20}99.99 & \cellcolor{Gray!20}96.01 & \cellcolor{Gray!20}91.79 & 79.84 & 60.38 & \cellcolor{Gray!20}93.22 & 83.52 & 72.71 & 64.08 & 55.95 & \cellcolor{Gray!20}98.31 & 72.95 & 45.80 & 32.22 & 24.46 \\
\textbf{E(50s)} & \cellcolor{Gray!20}99.99 & \cellcolor{Gray!20}97.76 & \cellcolor{Gray!20}97.31 & \cellcolor{Gray!20}95.67 & \cellcolor{Gray!20}93.00 & \cellcolor{Gray!20}95.81 & 88.58 & 81.13 & 73.96 & 68.23 & \cellcolor{Gray!20}99.18 & 89.28 & 64.86 & 47.77 & 37.46 \\
\bottomrule
\multicolumn{16}{r}{P=periodical, E=exponential}
\end{tabularx} \vspace{-3mm}
\label{tab:pdr_distr}
\end{center}
\end{table}
\begin{table}[!t]
\small
\setlength{\tabcolsep}{3pt}
\caption{E2E Delays (s) of \textit{Ctrl-MAC}/LoRaWAN/LoRaWAN++ under different data loads and data distributions.}
\vspace{-3mm}
\renewcommand{\arraystretch}{0.8}
\begin{center}
\begin{tabularx}{0.65\textwidth}{l | c c c c c | c c c c c}
\toprule
& \multicolumn{5}{c|}{\textbf{\textit{Ctrl}-MAC}} & \multicolumn{5}{c}{\textbf{LoRaWAN++}}\\
\midrule
\textbf{Size} & \textbf{10}& \textbf{50} & \textbf{100} & \textbf{150} & \textbf{200}  & \textbf{10} & \textbf{50} & \textbf{100} & \textbf{150} & \textbf{200}\\
\midrule
\textbf{P(10s)} & \cellcolor{Gray!20}1.38 & \cellcolor{Gray!20}9.17 & 23.37 & 37.66 & 52.77 & \cellcolor{Gray!20}4.15 & \cellcolor{Gray!20}14.39 & 15.68 & 15.56 & \cellcolor{Gray!20}14.84 \\
\textbf{P(30s)} & \cellcolor{Gray!20}1.28 & \cellcolor{Gray!20}1.48 & \cellcolor{Gray!20}5.85 & 26.31 & 36.02 & \cellcolor{Gray!20}3.32 & \cellcolor{Gray!20}11.92 & 15.26 & 15.60 & 15.40 \\
\textbf{P(50s)} & \cellcolor{Gray!20}1.27 & \cellcolor{Gray!20}1.41 & \cellcolor{Gray!20}1.66 & \cellcolor{Gray!20}2.8 & 25.54 & \cellcolor{Gray!20}0.33 & \cellcolor{Gray!20}8.54 & \cellcolor{Gray!20}14.52 & 15.57 & 15.26 \\
\textbf{E(10s)} & \cellcolor{Gray!20}1.35 & \cellcolor{Gray!20}2.81 & 17.29 & 31.96 & 46.65 & \cellcolor{Gray!20}3.58 & \cellcolor{Gray!20}13.41 & 15.52 & 15.63 & 15.03  \\
\textbf{E(30s)} & \cellcolor{Gray!20}1.30 & \cellcolor{Gray!20}1.48 & \cellcolor{Gray!20}2.91 & \cellcolor{Gray!20}14.99 & 27.03 & \cellcolor{Gray!20}1.63 & \cellcolor{Gray!20}9.51 & \cellcolor{Gray!20}14.33 & 15.56 & 15.38 \\
\textbf{E(50s)} & \cellcolor{Gray!20}1.29 & \cellcolor{Gray!20}1.37 & \cellcolor{Gray!20}1.59 & \cellcolor{Gray!20}2.45 & \cellcolor{Gray!20}7.73 & \cellcolor{Gray!20}1.16 & \cellcolor{Gray!20}6.36 & \cellcolor{Gray!20}12.28 & \cellcolor{Gray!20}14.62 & 15.46 \\
\midrule
\multicolumn{11}{l}{\textbf{LoRaWAN} E2E Delays are 0.15 seconds for all simulation scenarios.} \\
\bottomrule
\multicolumn{11}{r}{P=periodical, E=exponential} 
\end{tabularx} \vspace{-3mm}
\label{tab:delays_distr}
\end{center}
\end{table}
\begin{table}[!t]
\small
\setlength{\tabcolsep}{3pt}
\caption{UL Reliability (\%) of \textit{Ctrl-MAC}/LoRaWAN/LoRaWAN++ for $10$ and $200$ nodes }
\vspace{-3mm}
\renewcommand{\arraystretch}{0.8}
\begin{center}
\begin{tabularx}{0.9\textwidth}{l | c c | c c | c c | c c | c c | c c}
\toprule
 & \multicolumn{2}{c|}{\textbf{P(10s)}} & \multicolumn{2}{c|}{\textbf{P(30s)}} & \multicolumn{2}{c|}{\textbf{P(50s)}} & \multicolumn{2}{c|}{\textbf{E(10s)}} & \multicolumn{2}{c|}{\textbf{E(30s)}} & \multicolumn{2}{c}{\textbf{E(50s)}} \\
\midrule
\textbf{Protocol} & \textbf{10} & \textbf{200}  & \textbf{10} & \textbf{200}  & \textbf{10} & \textbf{200}  & \textbf{10} & \textbf{200}  & \textbf{10} & \textbf{200}  & \textbf{10} & \textbf{200}\\
\midrule
\textbf{\textbf{LoRaWAN}} & \multicolumn{12}{c}{UL Reliability is 0\% for all simulation scenarios.}\\
\midrule
{\textbf{LoRaWAN++}} & 98.76 & 34.62 & 99.92 & 37.81 & 99.99 & 39.41 & 99.46 & 35.36 & 99.96 & 41.91 & 99.99 & 50.81 \\
\midrule
\textbf{\textit{Ctrl}-MAC} & \multicolumn{12}{c}{UL Reliability is 100\% for all simulation scenarios.}\\
\bottomrule
\multicolumn{13}{r}{P=periodical, E=exponential}
\end{tabularx} \vspace{-4mm}
\label{tab:UL_rel}
\end{center}
\end{table}

The results clearly demonstrate that LoRaWAN is not suitable for control systems as it stands; it was not designed for this.
It achieves very low E2E PDRs (less than 90\%), except in the case of very small network sizes 
and low data rates. This implies failing
to support two-way communication. LoRaWAN does not support acknowledgement messages which prevents 
reliable data transmission to the gateway. The E2E Delays are constant and low as LoRaWAN uses pure ALOHA.
This implies that any contention-free transmission will have same E2E delays which would be a sum of the uplink and downlink time on airs which equals 0.15 seconds;
however, the percentages of contention-free transmissions are very low as per Table~\ref{tab:pdr_distr}.
Stable controlled system operation that we derive in Sec.~\ref{sec:stab_anal} can not be guaranteed
as it is shown in \cite{Tomic2018} with LoRaWAN.

Similar to LoRaWAN, LoRaWAN++ is also unable to meet the communication requirements for wide area control systems.
LoRaWAN++ achieves good E2E PDRs (more than 90\%) for small network sizes and low data rates.
However, for larger network sizes and higher data rates, there is a network performance
deterioration; E2E PDRs are even lower than in the case of LoRaWAN.
These results align with the results in \cite{Pop2018} which show that if 100\% of the nodes request
an acknowledgement, the network can barely operate at 15\% of its capacity compared to the scenario
when no-one requests an acknowledgement. LoRaWAN++ E2E Delays are always less than $20$ seconds as
it allows $8$ retransmissions, after which the packet is dropped. This means that LoRaWAN++ can provide
guarantees of bounded communication delays. However, it fails to guarantee reliable and two-way
communication. Not all delivered messages are acknowledged (based on the results in
Table~\ref{tab:UL_rel}) and this discrepancy increases with the number of nodes adding uncertainty to a closed-loop system.

Finally, the results show that \textit{Ctrl-MAC} is able to meet all communication requirements
for control systems for the networks of up to 150 nodes with a maximum data rate of 1 packet per 
50 seconds. Higher rates can be supported for the networks with less than 150 nodes (see 
Table~\ref{tab:pdr_distr}). This practical limitation of \textit{Ctrl-MAC} for networks larger
than 150 nodes per gateway is reasonable and as per general LoRa recommendations. To increase the scale
of such a single-hop network one would include more gateways to have gateway diversity rather than trying to handle more nodes per gateway. \textit{Ctrl-MAC} achieves high E2E PDRs for a much larger number of scenarios
when compared to LoRaWAN and LoRaWAN++. It achieves bounded E2E Delays and is also able to
guarantee UL reliability of 100\% for all simulated scenarios. This UL reliability can be guaranteed because all data transmissions are scheduled; hence, there is no possibility
of data collisions.

\section{Control Model for a Water Distribution Network} \label{sec:water_sys_control}

In this section, we present a system model for our large-infrastructure reference example,
a water distribution network. We demonstrate how we use our $C^{3}$ approach to influence the controller
design to meet the co-design constraint \textbf{C\textsubscript{2}} in Sec.~\ref{subseq:design_goals}
and enable safe and reliable system operation. The communication requirements of the controller
(based on the choice of $\{h_i, {\tau_d}_i, \sigma_i, \rho_i \}$) cannot be larger than the maximum
communication network capacity. At the same time, the parameters 
$\{h_i, {\tau_d}_i, \sigma_i, \rho_i \}$ have to be chosen such they guarantee the system
stability.

\subsection{System Model \& Controller Design}

\begin{figure}[!t]
\begin{center}
\includegraphics*[width=0.45\linewidth]{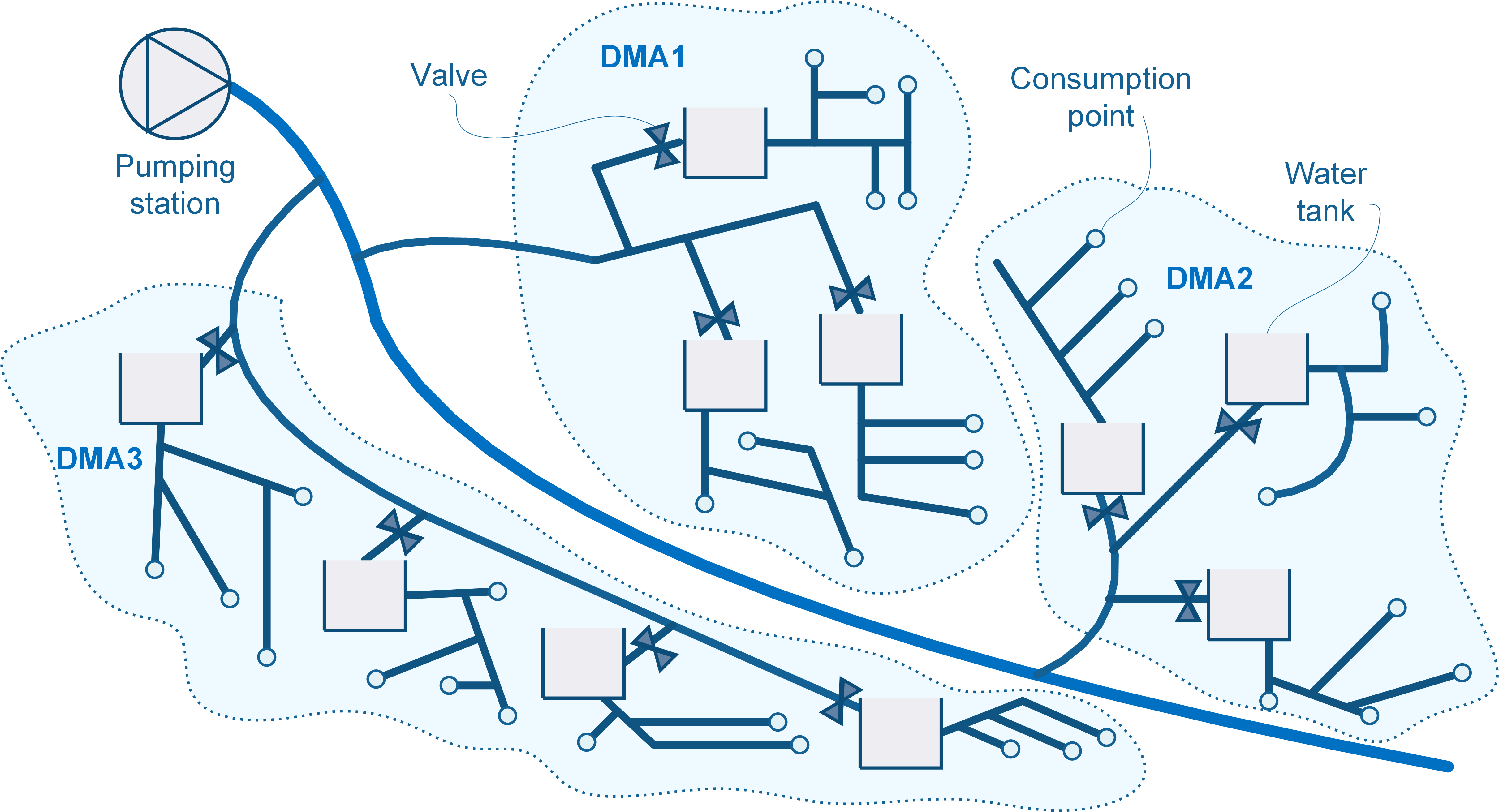}
\caption{\label{fig:water_net} The water distribution network.}
\vspace{-5mm}
\end{center}
\end{figure}

The water distribution network model considered in this paper is shown in
Fig.~\ref{fig:water_net}.
It consists of several District Meter Areas (DMAs). Each DMA has three or four water
tanks that supply on average $10$-$30$ customer connections via a set of pressurised
pipes, pumping stations, and valves. Their dynamics are complex, non-linear, difficult to model and analyse. If the complexity of the water network is reduced to
a problem of stabilising the tank water level to a pre-defined set point, the non-linear model can be linearised as in \cite{kartakis2017communication}.

Each DMA (or plant, in general terms) is instrumented with $n_i$ sensor nodes (for measuring
the water level in the tanks) and $m_i$ valves for actuation. Each DMA is modelled by
the linear time-invariant model in Eq.~\ref{dpetc:eq:plant}. In the state vector $\xi_i(t)$,
each state is the difference between the current water level and set point reference water
level for a tank. The control input vector $v_i(t)$ represents the degree to which the
in-valves are open to fill each tank.

The system states are measured with water level sensor nodes for each DMA supply tank. The
states are communicated to the linear input-to-state controller given in
Eq.~\ref{dpetc:eq:controller1}. The values in $\hat\xi_i(t)$ contain either the previous
measured values or new measured values of $\xi_i(t)$. A new measured value for a tank is
used if the measured value violates the event condition (defined in next section).
Otherwise, the previous measurement of the water level in the DMA tanks is used. The
combination of previous and new values creates an estimation of $\xi_i(t)$. The 
error, or difference, between the estimated levels and the measured levels is
\begin{equation}
\label{dpetc:eq:error}
\varepsilon_i(t)=\hat{\xi}_i(t)-\xi_i(t).  
\end{equation}
From a control system point of view, we can view the water tank levels in
$\xi_i(t)$ as those on the sensor nodes, and the values in $\hat{\xi}_i(t)$ as
the values on the controller.

It is the task of \textit{Ctrl-MAC} to send the new values from the sensors to the
controller when required. We refer to this process as the sample-and-update of
$\xi_i(t)$. The speed at which updates occur determines the system stability and the congestion of the communication network which relates to the co-design
constraint \textbf{C\textsubscript{2}}. The speed is regulated by the update
mechanism given next.

\subsection{Updating the Control Input}

The traditional way of choosing when to update $\hat{\xi}_i(t)$ is periodic,
or Time-Triggered Control (TTC). In TTC all of the sensors send updates
at a fixed interval. This method wastes energy when no changes to the water
levels warrant sending and can cause network congestion. Instead, we focus
on Decentralised Periodic Event-Triggered Control (DPETC) from
\cite{heemels2013periodic}. DPETC reduces both the communication from the
sensor nodes to the controllers and the energy consumption of the sensor nodes
\cite{kartakis2017communication}.

We summarise DPETC from \cite{heemels2013periodic} with modifications for
our input-to-state feedback control system (to accommodate the co-design
constraint \textbf{C\textsubscript{2}}). The equation 
\begin{equation}
\label{dpetc:eq:samplesequence}
\mathcal{T}_k:=\{t_k|t_k:=kh,k\in\mathbb{N}\}
\end{equation}
denotes the periodic sampling sequence of the sensor nodes in time. The term $h>0$ is
the time between samples and $k$ is the sample number. The sensor $s_{ij}$ of DMA $i$,
where $j \in \{1, \ldots, n_i\}$ updates its state, $\hat{\xi}_{ij}(t)$, on
the controller as
\begin{equation}
\label{dpetc:eq:update}
\hat{\xi}_{ij}(t)=\left\{
\begin{aligned}
&\xi_{ij}(t_k),\,\text{when }\mathcal{C}_{ij}(\xi_{ij}(t_k),\hat\xi_{ij}(t_k))>0\\
&\hat\xi_{ij}(t_k),\,\text{when }\mathcal{C}_{ij}(\xi_{ij}(t_k),\hat\xi_{ij}(t_k))\leq 0
\end{aligned}\right.
\end{equation}
for times $t\in(t_k,t_{k+1}]$. The quadratic event-trigger condition
$\mathcal{C}_{ij}(\xi_{ij}(t),\hat\xi_{ij}(t))$ defines when there is an event and
a sensor needs to send its value ($\xi_{ij}(t)$) to update $\hat\xi_{ij}(t)$.
The event condition $\mathcal{C}_{ij}$ is
\begin{equation}\label{dpetc:eq:eventcondition}
\mathcal{C}_{ij}(\xi_{ij}(t),\hat{\xi}_{ij}(t))=|\hat{\xi}_{ij}(t)-\xi_{ij}(t)|-\sigma_{ij}|\xi_{ij}(t)|.
\end{equation}

Based on Eq.~\ref{dpetc:eq:eventcondition}, small $\sigma_{ij}$ triggers new events faster 
which for us means that the load to our communication network will be higher. With this in mind, we choose to use a $\sigma_{ij}$ that triggers the least number of events but still guarantees 
stable system operation as per the co-design constraint \textbf{C\textsubscript{2}}. In doing so,
we perform stability analysis for our DPETC based on \cite{borgers2018riccati}. The goal of the
analysis is to define the maximum allowable delay $\tau_d$ and $\sigma_{ij}=\sigma$ for all $i$,
for which the control system is stable. The stability analysis is presented next.

For our stability analysis, we define $\mathcal{J}$ to be a set that contains
the states transmitted at time $t_k$ and $\mathcal{J}_c$ to be a set of the states not transmitted
at time $t_k$. We define $\Gamma_j$ as a diagonal matrix where the $j$th diagonal element equals
$1$ and the rest of diagonal elements equal $0$.

\subsection{Stability Analysis} \label{sec:stab_anal}

The following stability analysis determines the
maximum allowable delay $\tau_d$ for the chosen values of $h$, $\sigma$ and $\rho$ that 
renders the controlled system stable. Please note that we dropped denoting subsystems by 
$i$ for the sake of clarity. However, the analysis refers to an individual subsystem 
(or DMA) given in Eq.~\ref{dpetc:eq:plant} and its controller in Eq.~\ref{dpetc:eq:controller1}.

The condition for having a globally exponentially stable
(GES) system is that there exist scalars $c>0$ and $\rho>0$ such that for all
$\xi(0)\in\mathbb{R}^{n_s}$ for all $t\in\mathbb{R}^+$ the following holds:
$|\xi(t)|\leq ce^{-\rho t}|\xi(0)|$.
The stability proof follows from the Corollary III.3 and Theorem V.2 of \cite{heemels2013periodic}, and Theorem IV.1 of \cite{borgers2018riccati} and it is
briefly summarised in the following corollary.

\begin{corollary}\label{dpetc:corollary:stabilitywithdelay}
Consider plant in Eq.~\ref{dpetc:eq:plant}, controller in Eq.~\ref{dpetc:eq:controller1}, sampling sequence in Eq.~\ref{dpetc:eq:samplesequence}, event-triggered mechanism in Eq.~\ref{dpetc:eq:update}, and event-triggered condition in  Eq.~\ref{dpetc:eq:eventcondition}. Let $\tau_d$ be a given maximum allowable delay and $\rho$ be a given convergence rate. Assume there exist matrices $P_{0h}\succ0$ and $P_{1d}\succ0$, and scalars $\acute{\mu}_{\mathcal{J}}^j\geq0$, $\acute{\mu}_{\mathcal{J}_c}^j\geq0$, $\tilde{\mu}_{\mathcal{J}}^j\geq0$, $\tilde{\mu}_{\mathcal{J}_c}^j\geq0$, $\hat{\mu}_{\mathcal{J}}^j\geq0$, $\hat{\mu}_{\mathcal{J}_c}^j\geq0$, where $j\in\{1,\cdots,n_s\}$, such that
\begin{equation} \label{eq:lmi}
\begin{aligned}    
\begin{bmatrix}
e^{-2\rho\tau_d}P_{0h}+G_1 & \acute{J}_{\mathcal{J}}^{\mathrm{T}}e^{\bar{A}^{\mathrm{T}}\tau_d}P_{1d} \\
(\acute{J}_{\mathcal{J}}^{\mathrm{T}}e^{\bar{A}^{\mathrm{T}}\tau_d}P_{1d})^{\mathrm{T}} & P_{1d} \\
\end{bmatrix}&\succ0\\
\begin{bmatrix}
e^{-2\rho(h-\tau_d)}P_{1d}+G_2 & \tilde{J}_{\mathcal{J}}^{\mathrm{T}}e^{\bar{A}^{\mathrm{T}}(h-\tau_d)}P_{0h} \\
(\tilde{J}_{\mathcal{J}}^{\mathrm{T}}e^{\bar{A}^{\mathrm{T}}(h-\tau_d)}P_{0h})^{\mathrm{T}} & P_{0h} \\
\end{bmatrix}&\succ0\\
\end{aligned}
\end{equation}
where
$G_1=\sum_{j\in\{1,\cdots,n_s\}}(-\acute{\mu}_{\mathcal{J}}^j\acute{Q}_j+\acute{\mu}_{\mathcal{J}_c}^j\acute{Q}_j-\tilde{\mu}_{\mathcal{J}}^j\tilde{Q}_j+\tilde{\mu}_{\mathcal{J}_c}^j\tilde{Q}_j)$, 
$G_2=\sum_{j\in\{1,\cdots,n_s\}}(-\hat{\mu}_{\mathcal{J}}^j\hat{Q}_j+\hat{\mu}_{\mathcal{J}_c}^j\hat{Q}_j)$
and
\begin{equation*}
\small
\begin{aligned}
&\bar{A}=\begin{bmatrix}
A & BK & 0 \\
0 & 0 & 0 \\
0 & 0 & 0 \\
\end{bmatrix}
\acute{Q}_j=\begin{bmatrix}
(1-\sigma^2)\Gamma_j & 0 & -\Gamma_j \\
0 & 0 & 0 \\
-\Gamma_j & 0 & \Gamma_j \\
\end{bmatrix}\\
&\tilde{Q}_j=\begin{bmatrix}
(1-\sigma^2)\Gamma_j & -\Gamma_j & 0 \\
-\Gamma_j & \Gamma_j & 0 \\
0 & 0 & 0 \\
\end{bmatrix}
\hat{Q}_j=\begin{bmatrix}
0 & 0 & 0 \\
0 & \Gamma_j & -\Gamma_j \\
0 & -\Gamma_j & (1-\sigma^2)\Gamma_j \\
\end{bmatrix}\\
&\acute{J}_{\mathcal{J}}=\begin{bmatrix}
I & 0 & 0 \\
0 & I & 0 \\
\Gamma_{\mathcal{J}} & 0 & I-\Gamma_{\mathcal{J}} \\
\end{bmatrix}
\tilde{J}_{\mathcal{J}}=\begin{bmatrix}
I & 0 & 0 \\
0 & I-\Gamma_{\mathcal{J}} & \Gamma_{\mathcal{J}} \\
0 & 0 & I \\
\end{bmatrix}\\
\end{aligned}
\normalsize
\end{equation*}
Then, the system is GES with decay rate $\rho$.
\end{corollary}

The proof of Corollary~4.1 is given in Appendix~\ref{sec:proof}. Based on Corollary~\ref{dpetc:corollary:stabilitywithdelay} and its proof, the feasible solution of
Eq.~\ref{eq:lmi} for our water distribution network can be found when $h \in [1,50]$ seconds, 
$\tau_d \in [0,15]$ seconds and $\sigma \in [0.01, 0.3]$. We chose the set of parameters
that guarantees stable system operation and also meets the co-design constraint \textbf{C\textsubscript{2}}. It ensures that the network load is less than the maximum 
network capacity which we show next.

\subsection{System Operation} \label{subsec:control_operation}

\begin{figure}[!t]
\begin{center}
\includegraphics*[width=0.9\linewidth]{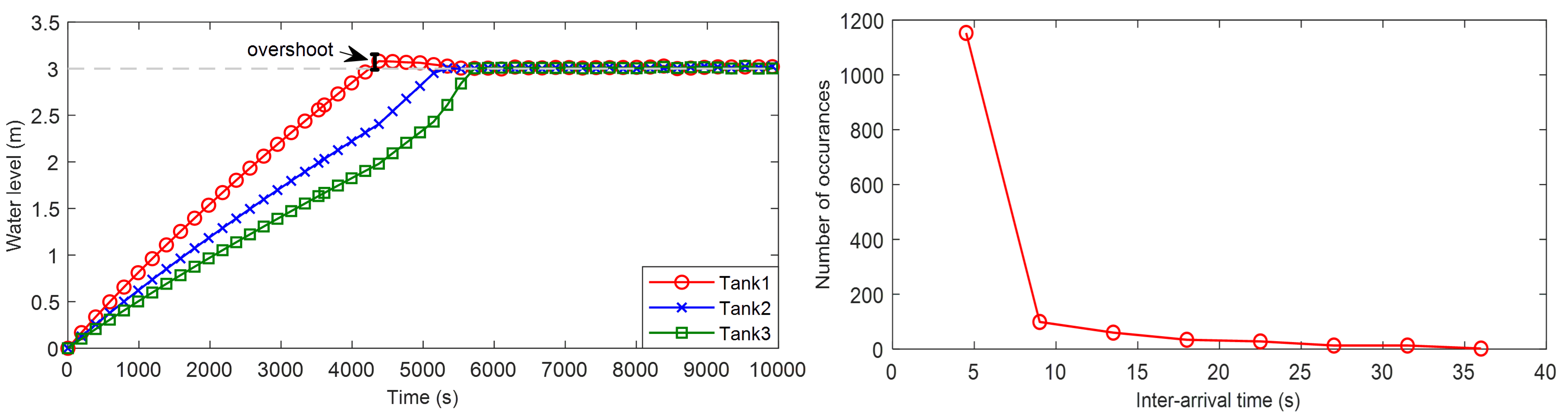}
\caption{\label{fig:system_response} (left) System response to constant customer demand, (right) Distribution of inter-arrival times of the events.}
\vspace{-3mm}
\end{center}
\end{figure}

We demonstrate the start and steady-state (or the system's response) of our model 
water distribution network in Fig.~\ref{fig:system_response} (left). 
The model is implemented and simulated using Matlab/Simulink, 
the Simscape toolbox and the Fluids toolbox. The parameters of the physical components
of the DMA (pipes diameters, valve types, tank sizes) are taken from the specification
sheets of the same make and model valves, pumps, and tanks as those used by water 
companies. 

We simulate an ideal case scenario with one DMA (with three tanks) and no packet drops or delays. The control system parameters we use are sampling time of $h=4.5$ seconds and the event trigger parameter of $\sigma=0.01$. $\sigma$ regulates the rate at
which events occur. For example, $\sigma=0.01$ leads to an average of $22.36$ events
per minute and increasing it to $0.1$ reduces the event rate to $13.87$ events
per minute. The event rate for both cases (which is the load to the communication network)
is much less than the network capacity of $136$ packets/minute which is needed to ensure
that the delays are bounded within a given threshold. This is presented in detail in Sec.~\ref{sec:delay}. Therefore, our control design meets the co-design constraint
\textbf{C\textsubscript{2}}.

We then plot the number of occurrences of the inter-arrival times as
shown in Fig.~\ref{fig:system_response} (right). By using the
curve-fitting toolbox,
this inter-arrival times can be approximated to an exponential
distribution with a
mean of $4.5$ seconds (equivalent to sampling time). This supports our
claim in choosing
the M/M/1 queue with exponentially distributed inter-arrival times to
model the input operation of \textit{Ctrl-MAC}. Additionally, multiple
exponentially distributed
event traces (as in the case of more than one DMA) models aperiodic 
and bursty traffic
\cite{jain2006} which is the reality of event-based systems.
In the next section, we evaluate the performance of our controller and 
communication protocol, or system, that we design using our $C^{3}$ 
approach.
\section{Evaluation} \label{sec:evaluation}

In this section, we evaluate a system designed using our $C^3$ approach that consists of a controller and communication protocol.  We refer to \textit{Ctrl-MAC} and its associated controller as the \textit{$C^3$ system}.
The evaluation is based on a co-simulation environment
and real-world experiments. We measure four system parameters to evaluate the
success of the $C^{3}$ system with respect to the co-design constraints defined in 
Sec.~\ref{subseq:design_goals}. The parameters are:
\begin{itemize}
    \item \textit{End-to-End Delay} (E2E Delay in seconds) - The time that starts when a sensor
    node senses an event and ends when the actuation update is successfully delivered to the actuator.
    This metric measures the success of the constraint \textbf{C\textsubscript{1}}.
    \item \textit{End-to-End Packet Delivery Ratio} (E2E PDR in \%) - The number of event messages received
    at the actuator over the number of messages generated by the sensor. For \textit{Ctrl-MAC} this metric measures the success
    of the contention resolution and successful actuation updates and this relates to the constraint \textbf{C\textsubscript{1}}.
    \item \textit{System overshoot percentage} (OvSh in \%) - The percentage of the overshoot above the
    reference value. It is calculated as the difference between the absolute value and reference value of
    the water level divided by the reference value. We use this metric to describe the stability of our
    system that has to hold in both constraints, \textbf{C\textsubscript{1}} and \textbf{C\textsubscript{2}}.
    \item \textit{Average number of events per minute} (Events/minute) - The average number of events per minute, measures the system load per minute. The more events, the more communication \textit{Ctrl-MAC} needs to handle. This metric measures the
    success of the constraint \textbf{C\textsubscript{2}}.
\end{itemize}

We measure these parameters for various sampling intervals, $h$, and different event
trigger parameters $\sigma$ as defined in Sec.~\ref{sec:sys_arch}. We vary the size of
the system, including different numbers of DMAs and sensor nodes. We conduct co-simulation and compare our results with real-world experiments to verify the success of the $C^3$ system.
In the following sections
we present our experiments and discuss the results.

\subsection{Evaluation via Co-Simulation}
\label{subsec:cosimeval}
We perform the first stage of the evaluation of the $C^{3}$ system on a custom co-simulation 
environment. This environment consists of a physical system simulator using 
MATLAB/Simulink and a communication simulator using the OMNeT++ network simulator. 
We run both Matlab and OMNeT++ on the same computer. We enable inter-process communication
between Matlab and OMNeT++ using UDP. 

We compare our approach to a 
wired centralised network and LoRaWAN++ in the co-simulation environment.
The wired network represents the best-case scenario of a periodic centralised
controller with no packet drops or transmission delays.
LoRaWAN++ uses the ALOHA contention scheme with acknowledgement messages and 
a retry limit of $8$ messages. 
This is the only configuration of LoRaWAN that can provide partial guarantees in message delivery required by control applications, as mentioned in Sec.~\ref{subsec_ctrlmac_lorawan}.
We use a spreading factor of $7$ and the default $4$ channels specified in the LoRaWAN specification.
We simulate the network with 3 DMAs and 10 sensor and actuator nodes.  
The goal of the control system is to maintain the DMA tank water levels 
to a reference point of $3$m as the time evolves. With the definition of overshoot percentage 
given above, a $4.5$m high water tank with a $3$m reference point overflows, or critically fails,
with an overshoot above $50\%$. The customer demand is 
expressed as the degree that the out-valve of the tank is open. The maximum is 
100\%, and the minimum, 0\%, is when the valve is closed.  \vspace{4pt}\\
\textbf{Study 1: Matching Constant Customer Demand.}
In this co-simulation, we keep the customer demand constant at 100\%.
All experiments were run for $9000$ seconds. 

\begin{figure}[!t]
\begin{center}
\includegraphics*[scale=.35]{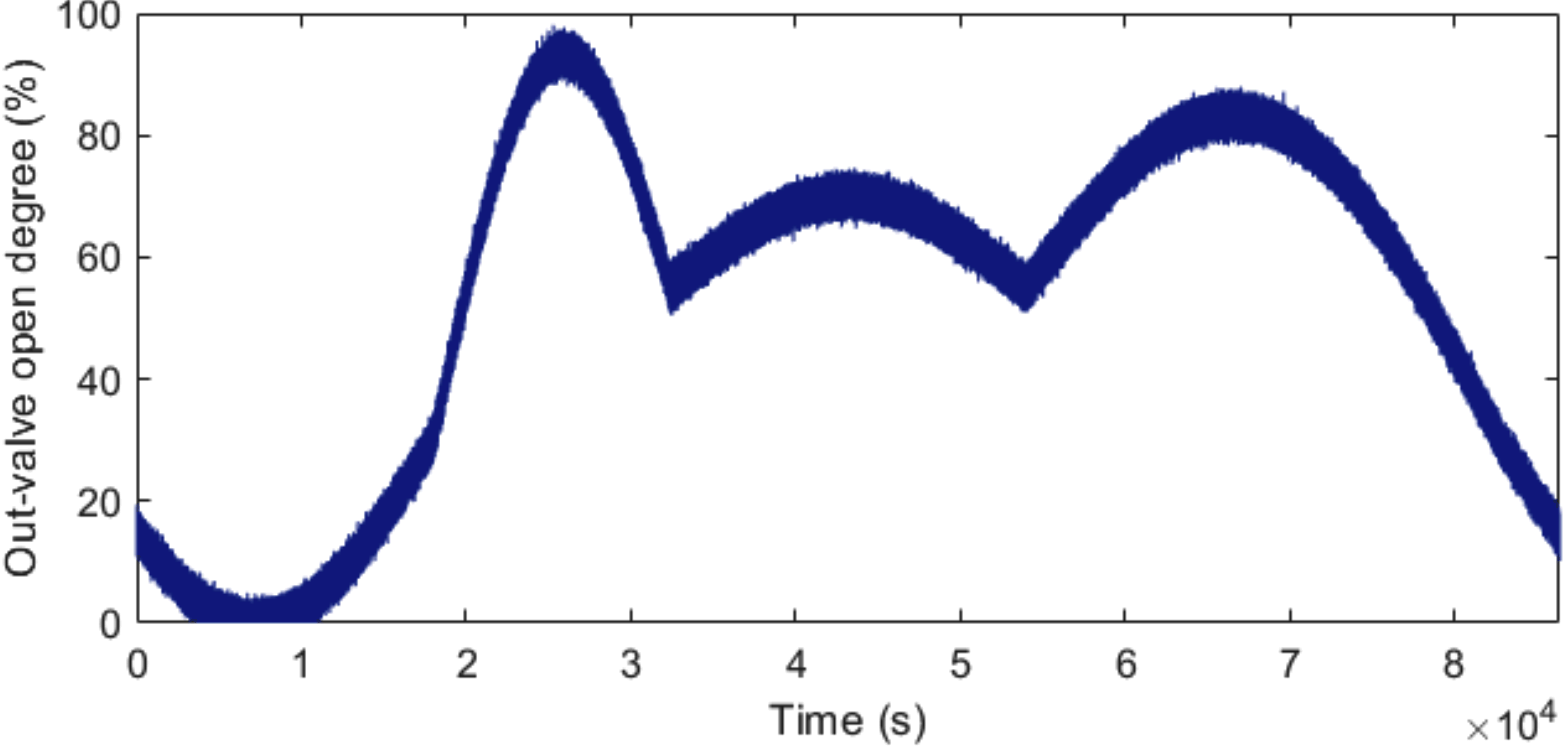}
\caption{\label{fig:demand} Daily customer demand ($24$ hours)}
\vspace{-4mm}
\end{center}
\end{figure}

The results in Table~\ref{tab:constant_dem} show that the $C^{3}$ system and LoRaWAN++ 
both maintain the stability of the system with similar overshoots. It is because
both \textit{Ctrl-MAC} and LoRaWAN++ always send the latest reading, and discard any messages that 
were not sent successfully. However, the E2E PDR of \textit{Ctrl-MAC} is on average 30\% higher, 
and more consistent than the E2E PDR of LoRaWAN++. 
The E2E Delays was bounded in all cases below 2.2 seconds for 
Ctrl-MAC. The LoRaWAN++ E2E Delays are on average greater than 6 seconds. 
With the $C^{3}$ system the events/minute are always less than $136$ packets/minute, indicating that the co-design constraint \textbf{C\textsubscript{2}} from
Sec.~\ref{subseq:design_goals} is being met, and the network load remains within
the network capacity.
\begin{table}[!t]
\small
\caption{The evaluation of the $C^{3}$ system for constant customer demand in comparison with an idealistic wired control and LoRaWAN++ for control.}
\vspace{-3mm}
\begin{center}
\renewcommand{\arraystretch}{0.8}
\begin{tabularx}{0.78\textwidth}{l c c c c}
\toprule
\textbf{$C^3$ system} & \textbf{Events/minute} & \textbf{OvSh (\%)} & \textbf{E2E PDR (\%)} & \textbf{E2E Delay (s)} \\ 
\midrule
$h=1$s, $\sigma=0.1$ & 113.15 & 20.59 & 77.33 & 1.849 \\
$h=4.5$s, $\sigma=0.01$& 61.17 & 32.11 & 87.03 & 2.161 \\
$h=4.5$s, $\sigma=0.1$& 40.32 & 3.1 & 88.74 & 2.194 \\
$h=4.5$s, $\sigma=0.3$& 22.56 & 4.39 & 88.01 & 1.861 \\
$h=10$s, $\sigma=0.1$& 14.91 & 16.25 & 89.27 & 1.919 \\
\midrule
\textbf{Wired Control} & \textbf{Events/minute} & \textbf{OvSh (\%)} & \textbf{E2E PDR (\%)} & \textbf{E2E Delay (s)} \\ 
\midrule
$h=-1$s* & $\inf$ & 2.73 & 100 & 0 \\
$h=1$s & 600 & 2.73 & 100 & 0 \\
$h=4.5$s & 133.33 & 2.75 & 100 & 0 \\
$h=10$s & 60 & 2.75 & 100 & 0 \\
\midrule
\textbf{LoRaWAN++} & \textbf{Events/minute} & \textbf{OvSh (\%)} & \textbf{E2E PDR (\%)} & \textbf{E2E Delay (s)} \\ 
\midrule
$h=1$s, $\sigma=0.1$ & 115.8 & 3.15 & 31.4 & 6.143 \\
$h=4.5$s, $\sigma=0.01$& 63.35 & 29.04 & 51.04 & 9.859 \\
$h=4.5$s, $\sigma=0.1$& 35.34 & 6.38 & 52.4 & 9.725 \\
$h=4.5$s, $\sigma=0.3$& 20.26 & 5.37 & 59.01 & 8.089 \\
$h=10$s, $\sigma=0.1$& 15.03 & 16.63 & 81.22 & 8.759 \\
\bottomrule
& & \multicolumn{3}{r}{*the system is continuous in time}
\end{tabularx} 
\vspace{-5mm}
\label{tab:constant_dem}
\end{center}
\end{table}
\vspace{4pt}\\
\textbf{Study 2: Matching Dynamical Customer Demand.}
In this co-simulation we model customer demand from actual traces of daily domestic water consumption inspired by \cite{gurung2015smart} and \cite{moreau2011control}. Fig.~\ref{fig:demand} shows the typical demand of a water network over 24 hours in an urban environment. It is a tri-modal curve in which the demands are highest during
the mornings and evenings and a small bump during the day time with the lowest demand at night.
This demand is simulated by controlling the degree to which the out-valve is open.

The results in Table~\ref{tab:dyn_demand} show that 
the $C^3$ system and LoRaWAN++ can both maintain systems stability through keeping small 
overshoots. The events per minute for the $C^3$ system are half of those for LoRaWAN++ with $\sigma=0.3$ and are within the maximum load that the system can handle as described in Sec.~\ref{sec:delay}. The $C^3$ system also has nearly 
twice the E2E PDR and 75\% smaller E2E Delays compared to LoRaWAN++. The $C^3$ system's E2E 
Delays are bounded by a maximum of 2.1 seconds. The E2E Delays for LoRaWAN++ are 8 seconds on average. The increase in E2E Delays and reduction of E2E PDR shows that a change in customer demand causes the 
LoRaWAN++ network load to approach its capacity and struggles  
to meet the co-design criteria \textbf{C\textsubscript{2}}.\vspace{4pt}\\
\begin{table}[!t]
\small
\caption{Evaluation of the $C^3$ system for dynamic customer demand
compared to idealistic wired control and LoRaWAN++ for control.}
\vspace{-3mm}
\begin{center}
\renewcommand{\arraystretch}{0.8}
\begin{tabularx}{0.78\textwidth}{l c c c c}
\toprule
\textbf{$C^3$ system} & \textbf{Events/minute} & \textbf{OvSh (\%)} & \textbf{E2E PDR (\%)} & \textbf{E2E Delays (s)} \\ 
\midrule
$h=4.5$s, $\sigma=0.1$& 50.33 & 6.67 & 87.3 & 2.043 \\
$h=4.5$s, $\sigma=0.3$& 34.80 & 6.0039 & 88.88 & 1.825 \\
\midrule
\textbf{Wired Control} & \textbf{Events/minute} & \textbf{OvSh (\%)} & \textbf{E2E PDR (\%)} & \textbf{E2E Delays (s)}  \\
\midrule
$h=4.5$s & 133.33 & 3.44 & 100 & 0 \\
\midrule
\textbf{LoRaWAN++} & \textbf{Events/minute} & \textbf{OvSh (\%)} & \textbf{E2E PDR (\%)} & \textbf{E2E Delays (s)}  \\
\midrule
$h=4.5$s, $\sigma=0.1$& 69.85 & 6.463 & 46.13 & 8.546 \\
$h=4.5$s, $\sigma=0.3$& 58.71 & 7.868 & 55.2 & 8.268 \\
\bottomrule
\end{tabularx} \vspace{-3mm}
\label{tab:dyn_demand}
\end{center}
\end{table}
\textbf{Study 3: Fault Detection.}
An important requirement of a water distribution network system is fault detection, when pipes 
burst due to sub-zero temperatures. A pipe burst causes 
water leakage which is another source of demand on a system. In this scenario, we assume that 
the sensors are able to take readings from the pipes 
(flow or vibrations) and determine if there is a leakage. 

The leak detection mechanism identifies a fault and communicates this to enable leakage mitigation. 
The controller closes the valves that provide water to the affected area, re-routing water and preventing waste. 
The authors of \cite{Kartakis2016} provide an example of this with leakage localisation to $50$cm \cite{yu2015}.

\begin{figure}[!t]
\begin{center}
\includegraphics*[width=0.58\linewidth]{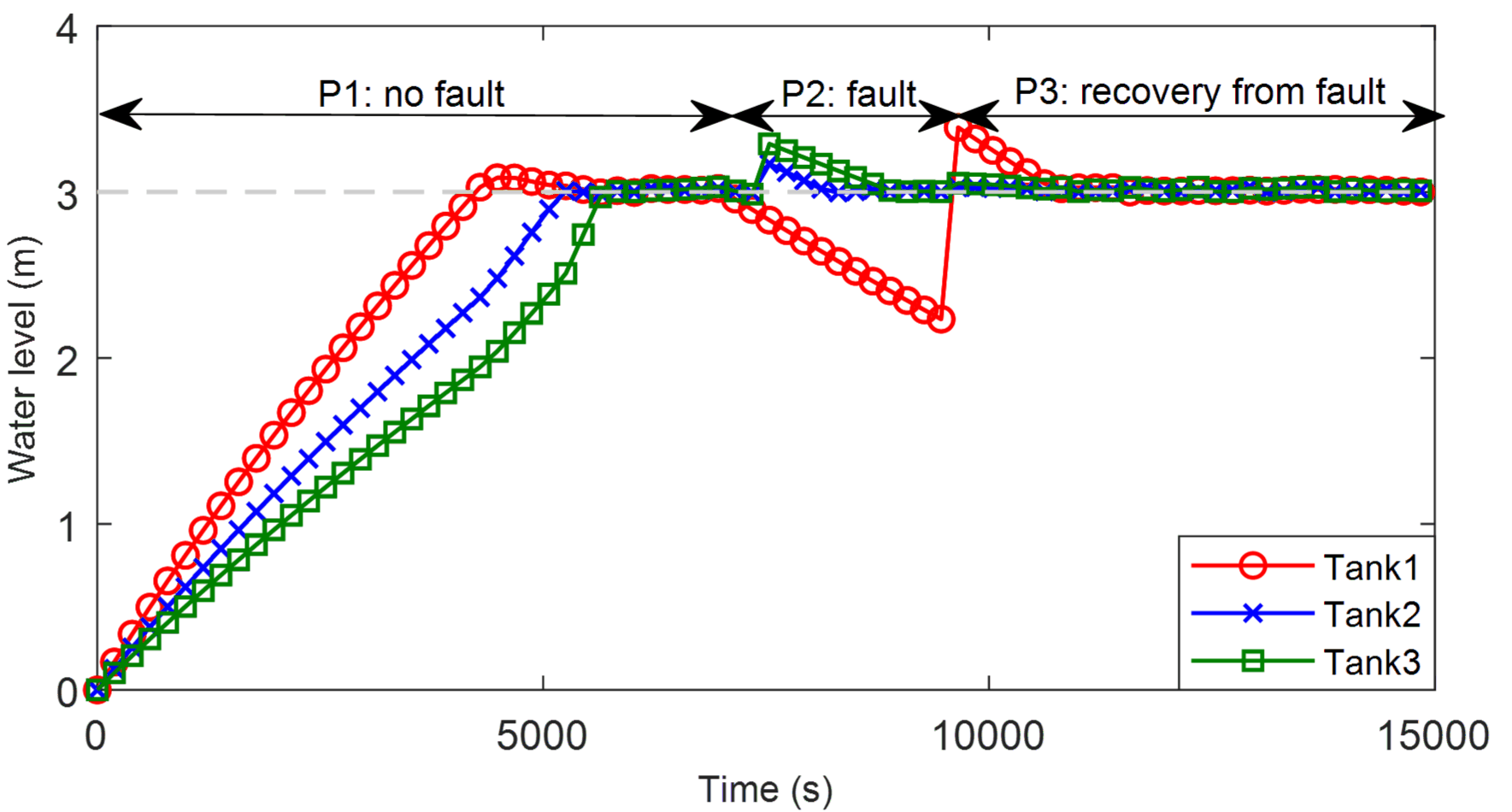}
\caption{\label{fig:system_response_fault} System response to a fault in the tank $1$.}
\vspace{-3mm}
\end{center}
\end{figure}

The system state response is depicted in Fig.~\ref{fig:system_response_fault} and has 
three phases: no-fault (P1), fault (P2) and recovery from fault (P3). We give
results in Table~\ref{tab:fault}.
The overshoots are comparable and stability is maintained. 
The E2E PDR of the $C^3$ system is almost twice that of LoRaWAN++. The E2E Delays are less than
2.1 seconds. The number of events/minute are within the maximum network load of
$136$ packets/minute and achieve high reliability.

\begin{table}[!t]
\small
\caption{The evaluation of the $C^3$ system for constant customer demand with fault in comparison with an idealistic wired control and LoRaWAN for control.\vspace{-3mm}}
\begin{center}
\renewcommand{\arraystretch}{0.8}
\begin{tabularx}{0.80\textwidth}{l l c c c}
\toprule
\textbf{$C^3$ system} & \textbf{Events/minute} & \textbf{OvSh (\%)} & \textbf{E2E PDR (\%)} & \textbf{E2E Delays (s)} \\ 
\midrule
\multirow{3}{*}{$h=4.5$s, $\sigma=0.1$} & \textbf{P1:} 25.03 & 3.3087  & \multirow{3}{*}{91.35} & \multirow{3}{*}{2.029} \\
& \textbf{P2:} 64.34 & 16.059  &  &  \\
& \textbf{P3:} 58.98 & 35.663  &  &  \\
\midrule
\multirow{3}{*}{$h=4.5$s, $\sigma=0.3$} & \textbf{P1:} 14.07 & 4.4741 & \multirow{3}{*}{92.25} & \multirow{3}{*}{1.789} \\
& \textbf{P2:} 41.87 & 6.3206 &  &  \\
& \textbf{P3:} 39.18 & 13.260 &  &  \\
\midrule
\textbf{Wired Control} & \textbf{Events/minute} & \textbf{OvSh (\%)} & \textbf{E2E PDR (\%)} & \textbf{E2E Delays (s)} \\ 
\midrule
\multirow{3}{*}{$h=4.5$s} & \textbf{P1:} 133.33 & 2.75 & \multirow{3}{*}{100} & \multirow{3}{*}{0} \\
& \textbf{P2:} 122.65 & 5.9129  &  &  \\
& \textbf{P3:} 133.32 & 10.23  &  &  \\
\midrule
\textbf{LoRaWAN++} & \textbf{Events/minute} & \textbf{OvSh (\%)} & \textbf{E2E PDR (\%)} & \textbf{E2E Delays (s)} \\ 
\midrule
\multirow{3}{*}{$h=4.5$s, $\sigma=0.1$}& \textbf{P1:} 25.52 & 3.0926 & \multirow{3}{*}{50.35} & \multirow{3}{*}{8.465} \\
& \textbf{P2:} 69.35 & 19.29 & & \\
& \textbf{P3:} 72.50 & 17.62 & & \\
\midrule
\multirow{3}{*}{$h=4.5$s, $\sigma=0.3$}& \textbf{P1:} 15.08 & 4.1367 & \multirow{3}{*}{60.94} & \multirow{3}{*}{6.875} \\
& \textbf{P2:} 38.25 & 7.4541 & & \\
& \textbf{P3:} 38.34 & 4.2372 & & \\
\bottomrule
& & \multicolumn{3}{r}{P1 - no fault, P2 - fault, P3 - recovery from fault}
\end{tabularx}
\vspace{-5mm}
\label{tab:fault}
\end{center}
\end{table}
The results in Table~\ref{tab:dyn_demand} show that LoRaWAN++
maintains system stability not by design but by accident. 
The E2E Delays of LoRaWAN++ are four times higher than those of
the $C^3$ system. Analysis of the simulation traces reveal that LoRaWAN++ succeeds in sending
uplink messages to the gateway, but gateway acknowledgements are lost due
to duty-cycle, $DC$, constraints.
The loss of acknowledgement messages causes a resend.  
The gateway forwards the received, unacknowledged, sensor value to the controller, regardless of whether
the sensor receives an acknowledgement. The sensor value leads to an actuation update and a stable system. The traces show that LoRaWAN++ is incapable of meeting
our co-design constraint \textbf{C\textsubscript{2}}, of maintaining the network
load within its capacity.
The freshness property of the sent data also enables LoRaWAN++ to maintain system stability. When a node fails to receive an acknowledgement before an event causes the sampling of new data, the new data replaces the old in the re-transmitted packet. In the case of lost acks and re-transmissions, the controller always receives the newest data and maintains system stability.
Our next study investigates whether the stability by accident of LoRaWAN++ scales to a large-scale system.
\vspace{4pt}\\
\textbf{Study 4: Large-Scale System.} In this co-simulation, we model a larger system with 10 DMAs 
consisting of 32 sensor and actuator nodes. We use constant customer demand, two sampling
intervals, $h$, of $1$ and $4.5$ seconds, and we run the experiment for $10000$ seconds.

The results in Table~\ref{tab:scale} show that at this scale LoRaWAN++ is unable to maintain
overshoots within the system boundaries and the controller is unable to maintain system stability. The $C^3$ system can maintain stability while achieving
overshoots of 22\% and a good performance in terms of E2E PDR and E2E Delays, which is consistent
with the three previous studies.
\begin{table}[!t]
\small
\caption{The evaluation the $C^3$ system for 10 DMAs (32 nodes) in
comparison with LoRaWAN++ for control.}
\vspace{-3mm}
\begin{center}
\renewcommand{\arraystretch}{0.8}
\begin{tabularx}{0.76\textwidth}{l c c c c}
\toprule
\textbf{$C^3$ System} & \textbf{Events/minute} & \textbf{OvSh (\%)} & \textbf{E2E PDR (\%)} & \textbf{E2E Delays (s)} \\ 
\midrule
$h=1$s & 253.93 & 32.44 & 63.35 & 3.36 \\
$h=4.5$s & 115.20 & 22.39 & 88.16 & 4.01 \\
\midrule
\textbf{LoRaWAN++} & \textbf{Events/minute} & \textbf{OvSh (\%)} & \textbf{E2E PDR (\%)} & \textbf{E2E Delays (s)} \\
\midrule
$h=1$s & 238.77 & 46.80 & 26.9 & 11.28  \\
$h=4.5$s & 174.26 & 63.03 & 45.2 & 14.9  \\
\bottomrule
\end{tabularx} \vspace{-5mm}
\label{tab:scale}
\end{center}
\end{table}

To conclude, our co-simulation results show that when the scale of the system increases by a 
factor of 3, LoRaWAN++ is no longer able to keep the network load within the capacity of the 
network and maintain system stability. This confirms our statement in 
Sec.~\ref{subsec_ctrlmac_lorawan} that the ALOHA channel access scheme is unable to provide the
communication requirements for control, and our motivation for the creation of the
two co-design constraints for our $C^{3}$ approach and the \textit{Ctrl-MAC}.

\subsection{Real World Evaluation} \label{subsec:realworldeval}

This section looks at the performance of the $C^3$ system 
designed using our $C^{3}$ approach
on real hardware.
For completeness, we apply a hybrid approach that combines a physical LoRa communication network
with a simulated model of a water supply network to create a hardware-in-the-loop experiment 
and a hardware-in-the-park experiment. This stage of the evaluation
validates the findings from the simulations showing that \textit{Ctrl-MAC} and the controller
can function correctly when tested with the uncertainty of real-world radio behaviours.

Real-world water distribution networks have mixed environments where they are deployed indoors and outdoors. We carry out hardware-in-the-loop experiment to characterise the behaviour of the $C^3$ system in an indoor environment. In this experiment, there is no line
of sight between nodes. They are separated by plaster, brick and metal walls at 
distances on the order of tens of meters. We divide ten sensor nodes into three groups, each representing a single DMA.

The hardware-in-the-park experiment is done in a $2\times2{\text{km}^2}$ urban park.
This experiment has buildings, trees and large ornamental structures
blocking line-of-sight. The distances of the hardware-in-the-park experiments are
in the order of hundreds of meters. In this experiment, 8 sensor nodes are divided up
into two DMAs.

\begin{figure}[!t]
\begin{center}
\includegraphics*[width=0.9\linewidth]{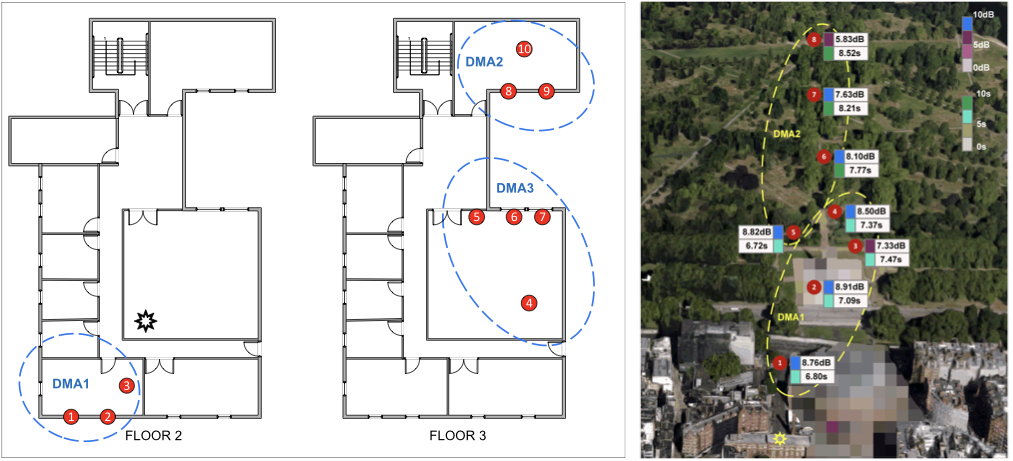}
\caption{\label{fig:hw_in_the_loop} (left) Hardware-in-the-Loop experimental setup, (right) Hardware-in-the-Park experimental setup with up-stream delay and SNR values as seen at the gateway  (the gateway is marked by the star, individual sensor nodes locations by red circles).}
\vspace{-3mm}
\end{center}
\end{figure}

\textit{Ctrl-MAC} is implemented on all sensor nodes. Each node consists of an
Adafruit Feather M0 RFM95 LoRa board mounted on a Raspberry Pi Model 3.
Both the Pi and the Feather are powered by large capacity batteries.
The gateway for the LoRa network is a MultiConnect Conduit 
Gateway connected to a desktop via Ethernet.
The gateway uses an 868MHz +3dbi whip antenna. We altered the packet forwarder 
software to provide a timestamp to upper layers of the gateway stack running on 
the desktop. The desktop runs Gateway-Bridge to receive communication from the gateway. The Gateway-Bridge then sends the messages to LoRa Server, 
for decryption. Following this, the LoRa Server sends the decrypted  messages
to an App Server. The entire software stack on the desktop is provided by ChripStack\footnote{\url{https://github.com/brocaar/chirpstack-network-server}}.
\vspace{4pt}\\
\textbf{Study 1: Hardware-in-the-Loop.}
Experimental setup is given in Fig.~\ref{fig:hw_in_the_loop} (left).
We provide the sensor nodes with a constant demand workload from
the physical system simulator as seen in the first co-simulation experiment.
We record the E2E Delays to measure the ability of our system to satisfy $\tau_d$,
and the overshoot to assess the ability of our approach to maintain system
stability.

The results are presented in Table.~\ref{tab:hw_in_loop}. There is a 
constant offset difference between the co-simulation delays and the 
experiment delays. These are due to issues with the hardware 
implementation at the gateway and unidentified latencies in the LoRa 
server software stack as the current LoRaWAN stack is not designed for
real-time operation. In spite of these added delays, the overshoots 
are still small and show that our $C^{3}$ approach can 
maintain system stability.\vspace{4pt}\\
\begin{table}[!t]
\small
\caption{Round Trip Times and Overshoot for Hardware-in-the-loop experiment.\vspace{-3mm}}
\begin{center}
\renewcommand{\arraystretch}{0.8}
\begin{tabularx}{0.86\textwidth}{l| c c c c c c c c c c}
\toprule
\textbf{Node ID} & \textbf{1} & \textbf{2} & \textbf{3} & \textbf{4} & \textbf{5} & \textbf{6} & \textbf{7} & \textbf{8} & \textbf{9} & \textbf{10}\\ 
\midrule
\textbf{Round Trip Time (s)} & 4.8 & 4.7 & 4 & 4.85 & 5.78 & 5.67 & 4.32 & 5.21 & 3.97 & 5.7 \\
\midrule
\textbf{Overshoot (\%)} & 4.98 & 17.25 & 3.8 & 8.4 & 12.36 & 7.22 & 9.64 & 4.98 & 16.6 & 13.3 \\
\bottomrule
\end{tabularx}
\vspace{-4mm}
\label{tab:hw_in_loop}
\end{center}
\end{table}
\textbf{Study 2: Hardware-in-the-Park.}
The experimental setup is shown in Fig.~\ref{fig:hw_in_the_loop} (right).
We chose this setup to provide realistic 
radio noise and effects such as the signal strength drop caused by distance and random
interference. These real-world disturbances are very difficult to categorise and model in 
simulation.
It is important to note that the small +3dbi whip antenna that we used on the gateway 
limited the range that we could achieve. The placement of the gateway, in the window of 
a building with line of site to the park, was also far from ideal and reduced our
overall range. 

This experiment differs from the previous hardware-in-the-loop because it does not use input directly from the simulator. Instead, this experiment uses event traces recorded from the simulator on every sensor node.
We measured the Round Trip Times (RTTs) and signal to noise ratio (SNR) at the gateway for each node in this 
experiment. RTTs are defined as the difference between when a packet was acknowledged and when it was generated.

The results are depicted in Fig.~\ref{fig:hw_in_the_loop} (right). It can be seen that the average
delay between two successful events is small, very similar to 
the previous hardware-in-the-loop experiments. The signal to noise ratios show that there 
is a correlation between distance, SNR and delays. This is an indication of higher packet 
loss and retransmissions. We see that, even in the presence of an unstable radio link
environment, the $C^3$ system keeps the delays bounded.
It is clear that the bounded delay times that we observed in both of our real world 
experiments validate the co-simulation results and that our $C^3$ approach is useful to design control and communication protocols that can maintain the stability of a WA-CPS.
\vspace{4pt}\\
\textbf{Experimental Observations.}
During our experimental evaluation with real sensor nodes, we encountered some strange 
packet loss behaviour that was the result of collisions and the capture effect \cite{Voigt:2017:MII:3108009.3108093}.
The capture effect occurs when two packets arrive at the same time and
there is a collision. One of the two packets is successfully decoded if the resulting
RSSI of the received packet is higher than the receiver noise floor after the RSSI of
the lost packet has been subtracted from it. The result is that the SNR value of the
decoded packet reduces. For experiments involving hardware, we used low SNR thresholds
to indicate the occurrence of collisions when the gateway had detected none. The
gateway would then communicate to the node that did not successfully send its packet
that it needed to restart the data sending process. We calculate SNRs at various
positions and derived a threshold SNR value of 7dB in the laboratory, and between
6dB and 8dB for sensors in the outdoor experiment.

\begin{figure}[!t]
\begin{center}
\includegraphics*[width=0.55\linewidth]{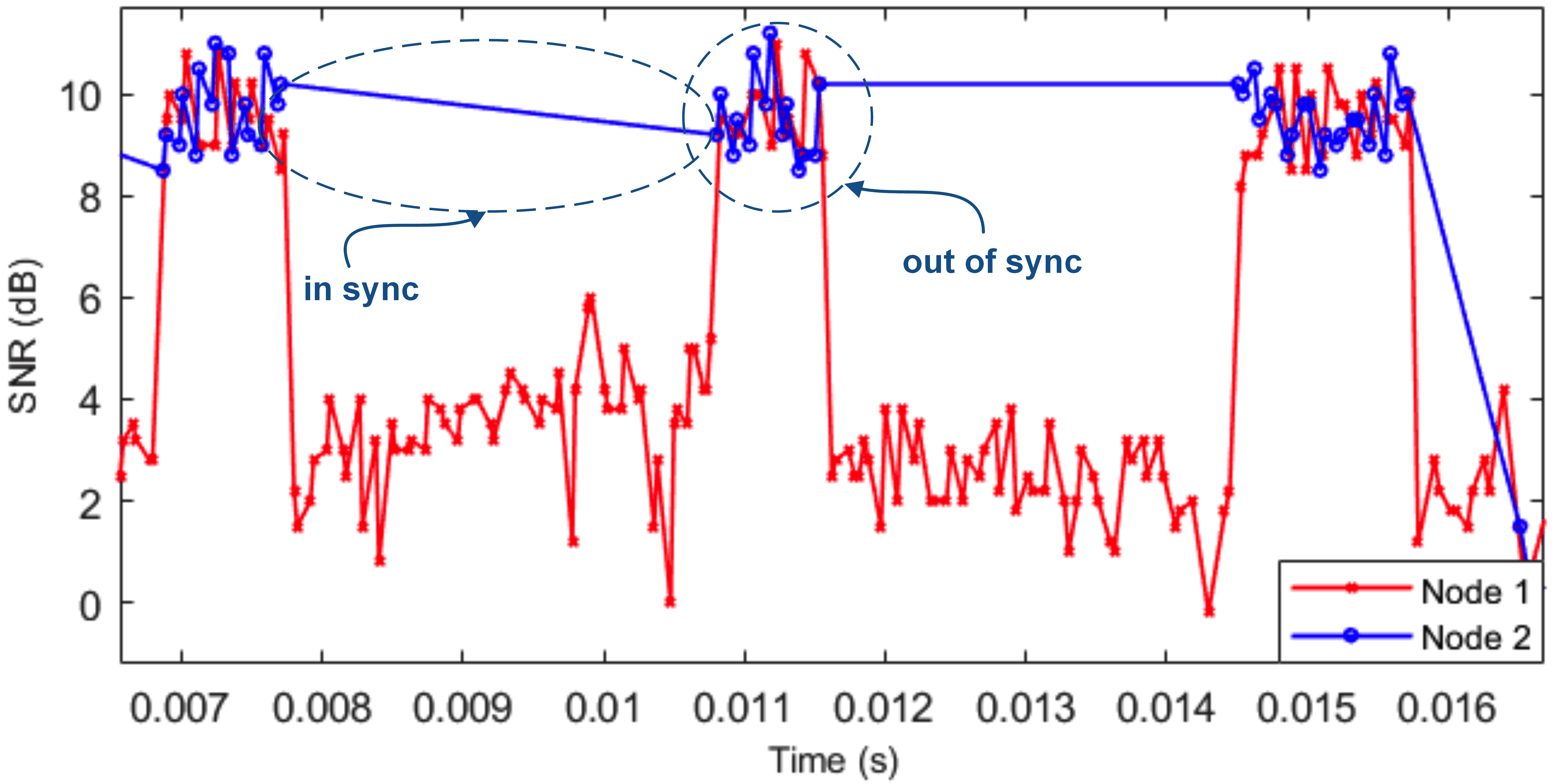}
\caption{\label{fig:capture} Capture effect phenomenon.}
\vspace{-3mm}
\end{center}
\end{figure}

The capture effect can be seen in Fig.~\ref{fig:capture} that shows two nodes that transmit at
the same time. The SNR for both nodes is between 9dB and 10dB for when
there are no collisions. During a collision with the capture effect, we see that the
SNR of the received packets is between 5dB and almost 0dB. We adapted \textit{Ctrl-MAC}
to use SNR threshold to detect collisions, which solved the problem.

\subsection{Evaluation Conclusion}

The results of our evaluation clearly show that the use of our $C^{3}$ approach enabled
the co-design of \textit{Ctrl-MAC} and a controller that can maintain the 
E2E communication delay bounds required to keep a WA-CPS physical system stable. 
The results in Table~\ref{tab:fault} show that LoRaWAN++ and the $C^3$ system can maintain
system stability on a 3 DMA system, although LoRaWAN++ has an E2E PDR of 50\% to 60\%
compared to 91-100\% for our $C^3$ system.
At this scale, LoRaWAN++ enables stability for small scale control systems by accident as described in Sec.~\ref{subsec:cosimeval}. 
As can be seen from Study 4, this situation does not scale well, and
LoRaWAN++ fails when the system scale increases. In Table~\ref{tab:scale}, we see that with 10 DMAs 
and 32 sensors LoRaWAN++ allows the overshoots to get as large as 63\%. The $C^{3}$ system keeps the overshoots down to 22\% for the same parameters. The E2E PDR falls to 26\% for LoRaWAN++, while the $C^3$ system has an E2E PDR of 63\% at a sample interval of 1 second.
Thus, the results of the evaluation show benefits of the $C^3$ system over LoRaWAN++.
\section{Related Work} \label{sec:related_work}
The contribution of this paper is in the combination of control and wireless computer communication. 
There is a large body of literature focusing on wireless network control systems both by co-designing the communication and control and considering them individually \cite{8166737}. We limit our review of the literature to event-triggered control and off-the-self wireless communication systems that have been used in such control systems. Similar approaches to the combination of control and communication are 
\cite{tubaishat2007adaptive,Suryadevara,pellegrino2016lighting,Liu2004,petersen2011,Eisen2019,Fadel,schindler2017implementation} but do not focus on event-triggered control.  

The most relevant work from the event-triggered control literature comes from \cite{Tabuada,mazo,Xi2019,lemmon}. In \cite{lemmon}, authors examine the bounds required to achieve event-triggered control. They use canonical control systems that do not reflect the complexity of real-world practical physical systems. Their work only addresses short-range communications (such as Zigbee) with high data rates and the inability to scale to multi-kilometres. In \cite{Tabuada}, the complexity of disturbances was ignored, and all sensors communicate local measurements synchronously. Both of these assumptions do not apply to large-scale physical systems. In \cite{lemmon}, the event trigger mechanism and controller are decentralised, making local computation complex (not suitable for low-powered sensor devices). In \cite{Xi2019} the authors propose real-time scheduling for event-triggered and time-triggered flows using 802.15.4 compliant radios. In \cite{mazo} the authors propose a decentralised event-triggered control with a TDMA protocol and reserved slots which will not scale to a large number of low-power nodes. 

The closest experimental works to ours from the event-triggered control literature includes \cite{araujo2014system,kartakis2017communication,mager}. In \cite{araujo2014system}, the system dynamics are entirely different from our own. The system has double tanks, a small area, uses Zigbee, and requires fast control of $20$~seconds to reach equilibrium. Our system has 10 tanks, covers a considerably larger area, uses LPWA, and is slow-loop requiring $4500$~seconds to reach equilibrium (see Fig.~\ref{fig:system_response}). The use of TDMA in \cite{araujo2014system} negatively impacts scale because nodes have to await their turn. Ctrl-MAC uses request slots to allow multiple nodes to request to send data immediately. The work in \cite{kartakis2017communication} considers centralised and decentralised event-triggered control of water systems. It also assumes LAN scale system on a small-scale emulator and not a distributed multi-subsystem. Exciting work by \cite{mager} derived communication schemes with guaranteed closed-loop stability for linear dynamic systems over low-power multi-hop networks. Their end-to-end design assumes a bespoke communication system based on constructive interference for control message flooding. The range of their solution is also limited to LAN scaled areas and not the wide-area multi-sub-system distributed control described in our paper.

\section{Limitations}
\label{sec:limitations}
While this paper aims to demonstrate that LPWA communication technologies can indeed support WA-CPS, there are some limitations to our exposition that we reserve for further research. 
Firstly, we do not explicitly discuss energy usage or saving. It is implicit in the way that we save large volumes of communication messages. However, we accept that actuation is relatively power-hungry (e.g. a pump) we either assume it is wired or uses energy harvesting and we have modelled the actuators as listening to the channel at all times, therefore we do not explicitly present energy costs of these in the results. Future work will look into minimising the listening time of the actuators to save energy. All other energy savings come from standard sensor communications approaches inherent in the lightweight LPWA protocol.
We scale our work to 150 nodes per base-station as per general LoRa recommendations. In real implementations of LoRa scale is achieved through the deployment of more gateways. However, this would expose different scheduling problems to manage feedback messages between many gateways to a given node; therefore, further work on slot allocation to cope with, or ensure, non-overlapping communications is required.
Robust analysis of a communications protocol in a control scenario dictates its analysis in terms of a canonical physical system, in our case, the water distribution network. This system was chosen as it represents the dynamics of many multi-kilometre pipeline systems,
further work on other applications will demonstrate generality further.
\section{Conclusion} \label{sec:conclusion}
In this paper, we present \textit{Ctrl-MAC}, a novel LPWA MAC protocol
that balances the control requirements of WA-CPS with the communication limitations imposed by LPWA communication technologies. 
Our protocol is a product of a Control Communication Co-design ($C^3$) approach for WA-CPSs.
The $C^3$ approach specifies a set of control and communication parameters and the constraints that couple them to create a controller and communication protocol that can maintain stability for a WA-CPS.
Our evaluation shows that Ctrl-MAC and its controller, referred to as the $C^3$ system, can maintain system stability under real-world conditions.
The results of our co-simulation evaluation show that neither LoRaWAN nor LoRaWAN++ can provide data reception guarantees to the sensors or the control system. 
Ctrl-MAC has up to $50$\% better average packet delivery ratio and $80$\% less average end-to-end delays when compared to LoRaWAN++. The combination of guarantees and low latencies makes Ctrl-MAC the only LPWA system suitable for the long range communication and control of WA-CPSs.

The evaluation criterion used in this paper aims to be acceptable to both the computer systems community (via reasoned algorithms, extensive scaled evaluations and real-world testing) and
the control community (via formal reasoning of a physical water distribution control system). 
Our evaluation shows that Ctrl-MAC in the $C^3$ system succeeds in maintaining stability based on the criterion that we establish for WA-CPS. This work is the first to demonstrate that LPWA communications, 
specifically (LoRa), can conclusively support a large class of control systems, previously considered beyond its scope. Our future work will continue to explore various applications of Ctrl-MAC in the wild.

\bibliographystyle{ACM-Reference-Format}
\bibliography{references}

\newpage
\appendix
\section{Appendix} \label{sec:app}
\subsection{Proof of Corollary \ref{dpetc:corollary:stabilitywithdelay}} \label{sec:proof}

    \begin{proof}
    The basic idea of the proof is inspired by the proofs of Theorem III.4 in \cite{borgers2018riccati} and Theorem III.2 and Corollary III.3 in \cite{heemels2013periodic}. The proof consists of three main steps which are presented next.
    
    In the first step, we consider the following Lyapunov function:
    \begin{equation}\label{eq:V}
    V(\xi,\tau)=\left\{
    \begin{aligned}
    &\xi^{\mathrm{T}}P_1(\tau)\xi,\,\tau\in[0,\tau_d]\\
    &\xi^{\mathrm{T}}P_0(\tau)\xi,\,\tau\in[\tau_d,h]\\
    \end{aligned}
    \right.
    \end{equation}
    where
    \begin{equation}\label{eq:P}
    \begin{aligned}
    \frac{\mathrm{d}}{\mathrm{d}\tau}P_i=-\bar{A}^{\mathrm{T}}P_i-P_i\bar{A}-2\rho P_i
    \end{aligned}
    \end{equation}
    with $i\in\{1,0\}$. The matrices $P_{0h}$ and $P_{1d}$ are defined in Corollary \ref{dpetc:corollary:stabilitywithdelay} as $P_0(h)=P_{0h}$ and $P_1(\tau_d)=P_{1d}$. Additionally, we define $P_{10}:=P_1(0)$, and $P_{0d}:=P_0(\tau_d)$.
    
    We also consider two scalars $c_1$ and $c_2$ that satisfy $0<c_1\leq c_2$ and are given as:
    \begin{equation*}
    \begin{aligned}
    c_1&:=\min\{\min_{\tau\in[0,\tau_d]}\lambda_{\min}(P_1(\tau)),\min_{\tau\in[0,h]}\lambda_{\min}(P_0(\tau))\}\\
    c_2&:=\max\{\max_{\tau\in[0,\tau_d]}\lambda_{\max}(P_1(\tau)),\max_{\tau\in[0,h]}\lambda_{\max}(P_0(\tau))\}\\
    \end{aligned}
    \end{equation*}
    where $\lambda_{\min}(P_i)$ and $\lambda_{\max}(P_i)$ are the minimum and maximum eigenvalues of matrix $P_i$, respectively. According to \cite{borgers2018riccati} and \cite{heemels2013periodic}, it can be show that the function $V$ satisfies the following inequalities:
    \begin{equation*}
    c_1|\xi|^2\leq V(\xi,\tau)\leq c_2|\xi|^2.
    \end{equation*}
    
    In the second step, we analyse $V$ during the flows and jumps of the system. Using Eq.~\ref{eq:V} and Eq.~\ref{eq:P}, and derivating $V$ along $\tau$ ($V$: $\tau\in[0,\tau_d]$ and $\tau\in[\tau_d,h]$) we get: 
    \begin{equation*}
    \frac{\mathrm{d}}{\mathrm{d}t}V(\xi,\tau)=-2\rho V(\xi,\tau).
    \end{equation*}
    
    In the third step, we need to show that $V$ does not increase during jumps. According to \cite{heemels2013periodic}, it holds that
    \begin{equation}\label{eq:prelation}
    \begin{aligned}
    P_{10}&=e^{2\rho\tau_d}e^{\bar{A}^{\mathrm{T}}\tau_d}P_{1d}e^{\bar{A}\tau_d}\\
    P_{0d}&=e^{2\rho(h-\tau_d)}e^{\bar{A}^{\mathrm{T}}(h-\tau_d)}P_{0h}e^{\bar{A}(h-\tau_d)}.
    \end{aligned}
    \end{equation}
    By applying Schur complement and S-procedure, it follows from Eq.~\ref{eq:lmi} that:
    \begin{equation*}
    \begin{aligned}
    &\xi^{+\mathrm{T}}P_{10}\xi^{+}=\xi^{\mathrm{T}}\acute{J}_{\mathcal{J}}^{\mathrm{T}}P_{10}\acute{J}_{\mathcal{J}}\xi\leq\xi^{\mathrm{T}}P_{0h}\xi\\
    &\xi^{+\mathrm{T}}P_{0d}\xi^{+}=\xi^{\mathrm{T}}\tilde{J}_{\mathcal{J}}^{\mathrm{T}}P_{0d}\tilde{J}_{\mathcal{J}}\xi\leq\xi^{\mathrm{T}}P_{1d}\xi.\\
    \end{aligned}
    \end{equation*}
    
    The results of this proof guarantee that $V$ is a proper Lyapunov function and the system is GES with convergence rate $\rho$ which concludes the proof.
    \end{proof}

\end{document}